\documentclass[useAMS,usenatbib,usegraphicx]{mn2e}
\usepackage{graphicx,times,epsfig}

\def\aap{A\&A}

\def\apjl{ApJ}
\def\apjs{ApJS}

\newcommand{\kmsend}{\mbox{km s$^{-1}$}}
 \newcommand{\kms}{\mbox{km s$^{-1}$ }}

\newcommand{\hpc}{\mbox{h$^{-1}$pc }}

\newcommand{\msun}{\mbox{M$_{\sun}$ }}
\newcommand{\msunend}{\mbox{M$_{\sun}$}}
\newcommand{\lsun}{\mbox{L$_{\sun}$ }}
\newcommand{\lsunend}{\mbox{L$_{\sun}$}}
\newcommand{\lir}{\mbox{L$_{\rm IR}$}}

\newcommand{\msunyr}{\mbox{M$_{\sun}$yr$^{-1}$ }}
\newcommand{\msunyrend}{\mbox{M$_{\sun}$yr$^{-1}$}}
\newcommand{\htwo}{\mbox{H$_2$}}

\newcommand{\z}{\mbox{$z$}}

\newcommand{\zsim}{\mbox{$z\sim$ }}

\newcommand{\magorrian}{\mbox{$M_{\rm BH}$-$M_{\rm bulge}$ }}

\newcommand{\sef}{\mbox{$F_{\rm 850}$}}
\newcommand{\stf}{\mbox{$F_{\rm 24}$}}
\newcommand{\sr}{\mbox{$F_{\rm R}$}}
\newcommand{\tfr}{\mbox{24/$R$}}
\newcommand{\microjy}{\mbox{$\mu$Jy}}
\newcommand{\sunrise}{\mbox{\sc sunrise}}
\newcommand{\gadget}{\mbox{\sc gadget-2}}
\newcommand{\starburst}{\mbox{\sc starburst99}}
\newcommand{\mappings}{\mbox{\sc mappingsiii}}
\newcommand{\turtlebeach}{\mbox{\sc turtlebeach}}
\newcommand{\bzk}{\mbox{{\it BzK}}}

\title[The Nature of \zsim 2 Dust-Obscured Galaxies]{A Physical Model
  for \zsim 2 Dust-Obscured Galaxies\thanks{This paper is dedicated
 to the original DOG himself, Cordozar Calvin Broadus, Jr.}}

\author[Desika and the Dog Pound]{Desika\,
  Narayanan$^{1}$\thanks{E-mail:
    dnarayanan@cfa.harvard.edu}\thanks{CfA Fellow}, Arjun Dey$^{2}$,
  Christopher C. Hayward$^1$, Thomas\, J.\,
  Cox$^{1,3}$\thanks{Carnegie Fellow}, \and R. Shane Bussmann$^4$,
  Mark Brodwin$^1$\thanks{W.M. Keck Postdoctoral Fellow}, Patrik\,
  Jonsson$^{5}$, Philip F. Hopkins$^6$\thanks{Miller Fellow}, \and
  Brent Groves$^7$, Joshua\, D.\, Younger$^{1,8}$\thanks{Hubble
    Fellow}, and Lars\, Hernquist$^{1}$\\$^{1}$Harvard-Smithsonian
  Center for Astrophysics, 60 Garden St., Cambridge, Ma
  02138\\$^2$National Optical Astronomy Observatory, 950 N. Cherry
  Avenue, Tucson, Arizona, 85719\\$^3$Observatories of the Carnegie
  Institute of Washington, 813 Santa Barbara St., Pasadena, Ca, 91101
  \\$^4$Steward Observatory, University of Arizona, 933 N Cherry
  Avenue, Tucson, Arizona, 85721\\$^{5}$Santa Cruz Institute for
  Particle Physics, University of California, Santa Cruz, Santa Cruz,
  Ca\\$^6$Department of Astronomy and Theoretical Astrophysics Center,
  University of California Berkeley, 601 Campbell Hall, Berkeley, Ca
  94720\\$^7$Sterrewacht Leiden, Leiden University, Neils Bohrweg 2,
  Leiden 233-CA, The Netherlands\\$^8$Institute for Advanced Study,
  Einstein Drive, Princeton, NJ, 08544}

\begin{document}

\date{Submitted to MNRAS}

\pagerange{\pageref{firstpage}--\pageref{lastpage}} \pubyear{2009}

\maketitle

\label{firstpage}

\begin{abstract}

We present a physical model for the origin of \zsim 2 Dust-Obscured
Galaxies (DOGs), a class of high-redshift ULIRGs selected at 24
\micron \ which are particularly optically faint ($F_{\rm 24\mu
  m}/F_R>1000$). By combining $N$-body/SPH simulations of high
redshift galaxy evolution with 3D polychromatic dust radiative
transfer models, we find that luminous DOGs (with \stf$\ga 0.3 $ mJy
at $z\sim 2$) are well-modeled as extreme gas-rich mergers in massive
($\sim 5\times10^{12}-10^{13}$ \msunend) halos, with elevated star
formation rates ($\sim 500-1000$ \msunyrend) and/or significant AGN
growth ($\dot{M}_{\rm BH} \ga 0.5$ \msunyrend), whereas less luminous
DOGs are more diverse in nature.  At final coalescence, merger-driven
DOGs transition from being starburst dominated to AGN dominated,
evolving from a ``bump'' to a power-law shaped mid-IR (IRAC) spectral
energy distribution (SED). After the DOG phase, the galaxy settles
back to exhibiting a ``bump'' SED with bluer colors and lower star
formation rates.  While canonically power-law galaxies are associated
with being AGN-dominated, we find that the power-law mid-IR SED can
owe both to direct AGN contribution, as well as to a heavily dust
obscured stellar bump at times that the galaxy is starburst
dominated. Thus power-law galaxies can be either starburst or AGN
dominated. Less luminous DOGs can be well-represented either by
mergers, or by massive ($M_{\rm baryon} \approx 5\times10^{11}$\msun)
secularly evolving gas-rich disc galaxies (with SFR $\ga 50$
\msunyrend).  By utilising similar models as those employed in the SMG
formation study of Narayanan et al. (2010), we investigate the
connection between DOGs and SMGs.  We find that the most heavily
star-forming merger driven DOGs can be selected as Submillimetre
Galaxies (SMGs), while both merger-driven and secularly evolving DOGs
typically satisfy the \bzk \ selection criteria.  The model SEDs from
the simulated galaxies match observed data reasonably well, though Mrk
231 and Arp 220 templates provide worse matches. Our models provide
testable predictions of the physical masses, dust temperatures, CO
line widths and location on the \magorrian \ relation of
DOGs. Finally, we provide public SED templates derived from these
simulations.

\end{abstract}

\begin{keywords}
cosmology:theory--galaxies:formation--galaxies:high-redshift--galaxies:starburst--galaxies:ISM--galaxies:ISM--ISM:dust
\end{keywords}

\section{Introduction}
\label{section:introduction}

Redshift \zsim 2 is a rich epoch for understanding galaxy formation
and evolution. During this time period, the bulk of the cosmic stellar
mass was assembled \citep{dic03,rud06}, and the star formation rate
and quasar space densities are both 
near their peaks \citep{bou04,hop04,hop07,sha96}.

In the local Universe, infrared (IR) selection of galaxies is an
efficient means of selecting galaxies undergoing significant star
formation (e.g., Arp 220), and possibly hosting optically-obscured
active galactic nuclei (AGN; e.g., Mrk 231, NGC 6240). By analogy, at
higher redshifts, IR selection techniques select galaxies which are
major contributors to the cosmic star formation rate density, far-IR
background, and progenitors of the present-day massive galaxy
population. Indeed, ultraluminous infrared galaxies (ULIRGs; \lir $>$
10$^{12}$ \lsunend) at \zsim 2 appear to contribute substantially to
the infrared luminosity density \citep[]{per05,cap07,red08,hop10,hop10b},
and the \z$>$0.7 star formation rate
density \citep{lef05,shi09}.

A well-studied sample of galaxies selected for their FIR properties at
\zsim 2 are Submillimetre Galaxies (SMGs), chosen in blank field
surveys for their redshifted cool dust emission at 850 \micron \ above
\sef $\ga$ 5 mJy. These galaxies are in a transient starburst phase
\citep[e.g.,] []{swi04,men07,you08b,mic09}, host rapidly growing
supermassive black holes \citep[e.g.,] []{ale05a,ale08}, and reside in
extremely massive $\sim$5$\times$10$^{12}$ \msun halos
\citep{bla04}. While these galaxies represent an important
sub-population of infrared-luminous galaxies at the epoch of peak
galaxy formation, there is some concern that a selection at 850
\micron \ may be missing significant populations of high-redshift
ULIRGs with warmer dust temperatures \citep[e.g.,]
[]{bus09,cha09,cas09,hua09,you09b}. As such, alternative selection
techniques for identifying \zsim 2 ULIRGs have become desirable.

With the launch of the {\it Spitzer Space Telescope}, surveys at 24
\micron \ have uncovered a population of ULIRGs at \zsim 2
\cite[e.g.,] [ and references
  therein]{rig04,don07,yan07,far08,soi08,lon09,hua09}.  Recently,
\citet[][ hereafter, D08; see also \citet{fio08}]{dey08} presented an
efficient means of identifying high-redshift ULIRGs which are both
mid-infrared luminous and optically faint. Specifically, by imposing a
nominal selection criteria of $R$-[24] $>$ 14 (roughly corresponding
to a flux density ratio \stf/\sr $>$ 1000; formally \stf/\sr $> 960$)
and \stf $>$ 0.3 mJy, D08 found a sample of ULIRGs with a relatively
narrow redshift distribution (centered at $<$\z$>$=1.99 with
$\sigma_z$=0.5) which exhibit heavy reddening of the intrinsic UV flux
and strong rest-frame 8 \micron \ emission \citep{hou05,fio08}. D08
refer to these galaxies as Dust-Obscured Galaxies (DOGs)\footnote{We
  note that while this paper nominally focuses on 24 \micron-selected
  galaxies which additionally have \tfr $> 1000$, the results of this
  paper are generally applicable to the 24 \micron-bright (\stf $> 0.3
  $mJy) \zsim 2 ULIRG population.}.

DOGs exhibit a range of luminosities, though AGN-dominated galaxies
tend to preferentially dominate the bright end of the 24 \micron
\ flux density distribution
\citep{bra06,wee06b,wee06a,dey08,fio08,fio09,sac09}. The DOGs with the
highest 24 \micron \ flux densities (\stf $>$ 0.8 mJy) have bolometric
infrared luminosities (\lir) $\sim$ 10$^{13}$ \lsun
\citep{bus09b,tyl09}.  As a population, the subset of \stf$>0.3$ mJy
galaxies which are also DOGs contribute around a quarter of the total
infrared luminosity density at \zsim 2 \citep{dey08}, and constitute
$\sim$60\% of the total contribution by \zsim 2 ULIRGs. It is clear that DOGs are a
cosmologically significant population of galaxies with diverse
observational characteristics.

Broadly, 24 \micron-selected ULIRGs at \zsim 2 (including DOGs) appear
to fall into two categories based on their mid-IR IRAC SEDs: those
with a power-law shape and those which exhibit a bump at observed
$\lambda \sim 5$ \micron \ corresponding to the 1.6 \micron \ rest
frame stellar photospheric bump (hereafter, these are referred to as
PL galaxies and bump galaxies, respectively). Bump galaxies are
thought to be dominated by star formation, with the bump (at
rest-frame 1.6 \micron) originating from starlight.  PL galaxies are
thought to have their mid-IR SED dominated by AGN continuum emission
\citep{wee06a,don07,yan07,mur09}.  In support of this scenario, PL
galaxies have SEDs which are reasonably represented by that of a Mrk
231 template \citep{bus09b,tyl09}, and tend to show relatively compact
optical morphologies \citep[$R_e\sim 1-6$ kpc;
][]{mel08,mel09,bus09}. On the other hand, bump galaxies are typically
polycyclic aromatic hydrocarbon (PAH)-rich, show PAH equivalent widths
consistent with being star formation dominated
\citep{yan05,saj07,far08,des09,hua09}, and are more often
morphologically resolved into multiple components
\citep{bus09}. However, \citet{fio08,fio09} have argued that even
these less luminous galaxies may harbor heavily obscured AGN.

DOGs are massive galaxies, with stellar masses $M_\star >$
10$^{10-11}$ \msun \citep{bus09,lon09,hua09}, and cluster in
group-sized halos of order \citep[$M_{\rm DM} \approx 10^{12-13}$
  \msunend;][]{bro08}. The clustering is luminosity-dependent, with
more luminous DOGs inhabiting more massive halos. Since the more
luminous DOGs also tend to exhibit power-law SEDs, it follows that PL
DOGs cluster more strongly than the lower luminosity bump DOGs.

Although much progress has been made in understanding DOGs in a
relatively short time period, a myriad of fundamental questions
regarding their physical nature exist. Are DOGs preferentially
isolated galaxies or mergers (or both)? Does the \tfr \ criterion
select galaxies at a particular evolutionary point? How are bump DOGs
and PL DOGs related - are they distinct galaxy populations, or
possibly related via an evolutionary sequence? How are DOGs related to
other coeval high-redshift populations (e.g., the SMGs, \bzk
\ galaxies, etc.)?

Numerical simulations can offer complementary information to the
observations in hand, and provide insight into the aforementioned
questions. In particular, by coupling radiative transfer modeling with
hydrodynamic simulations of galaxy evolution, one can precisely mimic
local and high-redshift observational selection functions, and
directly relate observed trends to physical conditions in the model
galaxies \citep[e.g.,] []{jon06a,jon06b,jon10,lot08,nar10a,you09}. 

Previously, we have utilised similar methods to investigate the
formation and evolution of high-redshift Submillimetre Galaxies
\citet[][C. Hayward et al. in prep.]{nar09b,nar10a}. Here, we build
upon these efforts by utilising similar models as those
employed in \citet{nar09b} and \citet{nar10a}.  This will allow us not
only to investigate a potential formation mechanism for DOGs, but to
explore the potential connection between \zsim 2 ULIRGs selected for
their 24 \micron \ brightness, and those selected in the
submillimetre.

The paper is organised as follows. In
\S~\ref{section:methods}, we detail our numerical methods. In
\S~\ref{section:dogformation}, we describe the formation of \zsim 2
DOGs, and explain the evolution of the 24/R ratio and the relationship
between bump and PL DOGS. In \S~\ref{section:physicalnature}, we
discuss the physical requirements necessary to form a DOG, and whether
mergers are necessary; In \S~\ref{section:othergalaxies}, we detail
the relationship between DOGs, SMGs and \bzk \ galaxies. In
\S~\ref{section:observations}, we relate our model to existing
observations, and make testable predictions. In
\S~\ref{section:discussion}, we provide discussion. We conclude in 
\S~\ref{section:conclusions}. 

Throughout this paper, we utilise a fiducial DOG selection criteria
\tfr \ $>$ 1000 and \stf $>$ 300 \microjy, consistent with the DOGs
sample in D08. We note again that while we focus on these 24 \micron
\ selected galaxies which are optically faint (DOGs), the results are
generally applicable to most 24 \micron -selected \zsim 2 ULIRGs.
Throughout, we assume a cosmology with $\Omega_\Lambda=0.7$,
$\Omega_m=0.3$ and $h=0.7$.  Finally, we note that the models
presented here are not cosmological in nature. As such, all
non-evolutionary plots should be taken to represent ranges of expected
values and colors, rather than true expected distributions.

\begin{table*}
\label{table:ICs}
\centering
\begin{minipage}{100mm}
\caption{DOGs are ordered with decreasing halo mass (and grouped by
  mergers and isolated galaxies, such that the isolated galaxies have
  the prefix 'i' in their name). Column 1 is the name of the model
  used in this work.  We emphasise that these models are nearly
  identical to those employed in the SMG formation study by
  \citet{nar10a} and \citet{nar09b}.  This will facilitate the
  comparison of these models with SMGs (\S~\ref{section:smg}). Columns
  2 and 3 give the virial velocity and halo mass of the
  galaxies. Column 4 is the total baryonic mass of the system. Column
  5 is the merger mass ratio. Columns 6 \& 7 are initial orientations
  for disc 1, Columns 8 \& 9 are for disc 2. The orientation angles
  are with respect to the merger plane (such that $\theta , \phi = 0$
  would be a coplanar merger). Column 10 lists the initial gas fraction
  of the simulation.}
\begin{tabular}{@{}cccccccccc@{}}
\hline Model & V$_{\rm c}$ & $M_{\rm DM}$ & $M_{\rm bar}$& Mass Ratio & $\theta_1$&$\phi_1$ & $\theta_2$ & $\phi_2$ &$f_g$\\ 
&(\kmsend)&\msun&\msun&&&&\\ \hline
DOG1 & 500:500 & 1.25$\times$10$^{13}$:1.25$\times$10$^{13}$ & 1.1$\times$10$^{12}$& 1:1 & 30 & 60 & -30 & 45 &0.8,0.6,0.4\\
DOG2 & 500:500 &  1.25$\times$10$^{13}$:1.25$\times$10$^{13}$ & 1.1$\times$10$^{12}$& 1:1& 360 & 60 & 150 & 0& 0.8\\
DOG3 & 500:500 &  1.25$\times$10$^{13}$:1.25$\times$10$^{13}$& 1.1$\times$10$^{12}$& 1:1 &-109 & -30 & 71 &-30& 0.8\\ 
DOG4 & 500:320 & 1.25$\times$10$^{13}$:3.4$\times$10$^{12}$ & 6.9$\times$10$^{11}$&1:3 & 30 & 60 & -30 & 45& 0.8,0.6,0.4\\ 
DOG5 & 500:320 &  1.25$\times$10$^{13}$:3.4$\times$10$^{12}$& 6.9$\times$10$^{11}$&1:3 & 360 & 60 & 150 & 0 & 0.8\\ 
DOG6 & 500:320 &  1.25$\times$10$^{13}$:3.4$\times$10$^{12}$& 6.9$\times$10$^{11}$&1:3& -109 & -30 & 71 & -30 & 0.8\\ 
DOG7 & 500:225 & 1.25$\times$10$^{13}$:1.2$\times$10$^{12}$ & 6.0$\times$10$^{11}$&1:10 & 30 & 60 & -30 &45 & 0.8,0.6,0.4\\ 
DOG8 & 500:225 & 1.25$\times$10$^{13}$:1.2$\times$10$^{12}$ & 6.0$\times$10$^{11}$&1:10 & 360 & 60 & 150 & 0 & 0.8\\ 
DOG9 & 500:225 & 1.25$\times$10$^{13}$:1.2$\times$10$^{12}$ & 6.0$\times$10$^{11}$&1:10 & -109 & -30 & 71 &-30 & 0.8\\ 
DOG10 & 320:320 & 3.4$\times$10$^{12}$:3.4$\times$10$^{12}$ &2.9$\times$10$^{11}$ & 1:1 & 30 & 60 & -30 &45 &0.8,0.6,0.4\\ 
DOG11 & 320:320 & 3.4$\times$10$^{12}$:3.4$\times$10$^{12}$ &2.9$\times$10$^{11}$ & 1:1 & 360 & 60 & 150 &0 &0.8\\ 
DOG12 & 320:320 & 3.4$\times$10$^{12}$:3.4$\times$10$^{12}$ &2.9$\times$10$^{11}$ & 1:1 &-109 & -30 & 71 &-30& 0.8\\ 
DOG13 & 225:225 & 1.2$\times$10$^{12}$:1.2$\times$10$^{12}$ & 1.0$\times$10$^{11}$& 1:1 & 30 & 60 & -30 &45 &0.8,0.6,0.4\\ 
iDOG1 & 500 & 1.25$\times$10$^{13}$ &5.5$\times$10$^{11}$& --- & N/A & N/A&N/A & N/A& 0.8,0.6,0.4\\ 
iDOG2 & 320 & 3.4$\times$10$^{12}$ & 1.5$\times$10$^{11}$& --- & N/A & N/A&N/A & N/A &0.8,0.6,0.4\\ 
iDOG3 & 225 & 1.2$\times$10$^{12}$ &5$\times$10$^{10}$& ---& N/A &N/A& N/A & N/A &0.8,0.6,0.4\\

\hline
\end{tabular}
\end{minipage}
\end{table*}

\section{Numerical Methods}
\label{section:methods}

Our methodology is as follows.  We first run a suite of
Smoothed-particle hydrodynamics (SPH) simulations of binary gas-rich
galaxy mergers over a range of baryonic masses and merger mass ratios.
The mergers are between idealised disc galaxies.  We additionally run
three isolated disc galaxy simulations as a control.  These
simulations represent potential physical models for \zsim 2 24
\micron\ sources.  We then investigate the UV-mm wave SEDs of these
galaxies utilising a 3D dust radiative transfer code in
post-processing.  In what follows, we describe the details of these
numerical simulations, and outline both the input parameters, as well
as the assumptions made.

\subsection{Hydrodynamics}
\label{section:hydro}

The hydrodynamic evolution of the galaxies was simulated utilising a
modified version of the publicly available $N$-body/SPH code, \gadget
\ \citep{spr05b}. \gadget \ follows the dynamical evolution of the dark
matter, ISM, stars and black holes in the simulated galaxies, and
explicitly conserves both energy and entropy \citep{spr02}.

The ISM is considered to comprise of a two-phase medium via a sub-grid
model \citep{spr03a}.  In this model, cold clouds are assumed to be
embedded in a hotter, pressure-confining phase. The cold phase may
convert to the hotter phase via evaporation from supernovae heating,
and similarly the hot phase may cool into cold gas.  This ISM is
pressurised by heating from supernovae.  Numerically, this is handled
via an effective equation of state \citep[hereafter EOS; see
  Figure 4 of ][]{spr05a}.  Here, we assume the stiffest equation of
state which stabilises the disc against runaway fragmentation.  We
explore the consequences of relaxing this assumption in
\S~\ref{section:freeparameters}.

The star formation rate in the simulations is dependent on the density
of cold gas (defined here as gas which has reached the floor of the cooling
curve, $T=10^4 K$) as SFR$ \propto \rho_{\rm gas}^{1.5}$.  The
normalisation of this relation is set such that the simulated star
formation rate surface density scales with the gas surface density in
agreement with local observations \citep{ken98b,cox06a}.

Black holes are included in the simulations as collisionless
particles.  These particles stochastically accrete matter according to
a Bondi-Hoyle-Lyttleton paramaterisation when a particle comes within
a smoothing length of the black hole sink particle.  To model the
impact of thermal feedback from accreting black holes, a fraction of
the accreted mass energy (here, 0.5\%) is reinjected thermally and
spherically into the surrounding ISM \citep{spr05a}. The fraction of
accreted mass energy which couples back to the ISM is set by matching
the normalisation to the local $M_{\rm BH}-\sigma$ relation \citep[though
  note that the slope is a natural consequence of the AGN feedback;
][]{dim05}.

The progenitor disc galaxies are initialised with a \citet{her90} dark
matter halo profile. The choice of concentration and virial radius for
a halo of a given mass was motivated by cosmological $N$-body
simulations, then scaled to match the expected redshift evolution
\citep{bul01,rob06b}. The halo for our fiducial model (DOG10,
Table~\ref{table:ICs}) has a \z=0 concentration $c=11.0$ and spin
parameter $\lambda = 0.033$.  We initialise our discs to reside at
\z=3, so that our simulated mergers occur around \zsim 2. We consider
initial virial velocities in the range 225-500 \kmsend, resulting in
dark matter halo masses of $\sim10^{12}$ \msunend-10$^{13}$\msunend,
consistent with the DOG clustering measurements by \citet{bro08}. The
halos are populated with exponential discs whose disc scale lengths
follow the \citet{mo98} prescriptions.

The discs for model galaxies are initialised with gas fractions
$f_{\rm g} \equiv {\rm gas/(gas+stars)} = 0.8$. Simulation tests by
\citet{nar10a} have shown that high-redshift mergers initialised with
this gas fraction result in $f_{\rm g} \approx0.2-0.4$ at the time
when the galaxy may be viewed as an SMG, which is comparable to
observations \citep[e.g ][]{bou07,tac08,dad10a,tac10}. While gas
fractions in DOGs are observationally unconstrained, based on an
observed overlap between the luminous DOG and SMG populations
\citep[e.g., ][]{pop08,bro08}, we consider this assumption
reasonable. In order to investigate any potential dependence on gas
fraction, we have run a subset of our simulations with initial gas
fractions $f_g = 0.4, 0.6$ as well.  The results from all
initialisations are qualitatively similar, and we include both in our
plots to increase the simulation sample size. The gravitational
softening lengths are set at 100 \hpc \ for baryons and 200 \hpc \ for
dark matter.  Because structure in the ISM clearly exists below these
scale lengths, we employ a subgrid specification to the ISM for the
radiative transfer calculations.  This will be discussed in the
following two subsections.

In this paper, we consider 16 simulations, with inital conditions
listed in Table~\ref{table:ICs}. The simulations (and nomenclature)
here are nearly identical\footnote{The primary difference between the
  simulations run here and those employed in \citet{nar09b} is in the
  assumptions of the evolving dust mass (see the next section).  Here, we tie
  the dust mass to the metallicity, whereas in \citet{nar09b}, the
  dust mass is tied to the gas mass.  This makes little difference
  during the merger event, though can have some effects on the optical
  flux during the inspiral phase.} to those in \citet{nar09b} and
\citet{nar10a}'s studies of SMG formation.  This will facilitate
direct comparisons between the DOG population and high-redshift SMGs.
When we refer to investigating a fiducial merger, this will refer to
model DOG10 with initial $f_g=0.8$.

\subsection{Radiative Transfer}
\label{section:sunrise}

\subsubsection{General Methods}

We extract synthetic SEDs from our SPH simulations in post-processing
utilising the 3D polychromatic Monte Carlo radiative transfer code,
\sunrise \ \citep{jon06a,jon10}. \sunrise \ considers the propagation
of UV through mm-wave photons through a dusty medium, and operates on
an adaptive mesh. Here, we summarise the aspects of the code most
relevant to this study, and refer the reader to \citet{jon10} for the
most current detailed description of the underlying algorithm.

Dust radiative transfer may be thought of in terms of sources and
sinks.  Here, the sources of radiation are stellar clusters, AGN and
dust grains.  The stellar clusters emit a spectrum calculated by
\starburst \ \citep{lei99,vaz05}, where the ages and metallicities of
the stars are known from the SPH simulations.  The stellar particles
present in the simulation's initial conditions are assumed to have
formed over a constant star formation history for $\sim$250 Myr (where
this time was chosen by dividing the stellar mass of the first
simulation snapshot by the SFR at that time). That said, the
rest-frame UV and mid-IR (which dominates the wavelength regime of
interest for DOGs) is contributed mainly by young ($< 100 $ Myr) stars
(and their surrounding HII regions), and thus our results are largely
insensitive to the choice of star formation history.  We assume a
Kroupa IMF for these calculations, suggested by
observations of \zsim 2 SMGs \citep{tac08}.

The AGN input spectrum utilises the luminosity-dependent templates of
\citet{hop07} of type I quasars. The template spectrum includes
torus hot dust emission, and is based on the mid-IR SED template of
\citet{ric06}. The normalisation of the input
spectrum is set by the total bolometric luminosity of the black
hole(s), which is determined by their mass accretion rate: $L_{\rm
  AGN} = \eta {\dot{M}_{\rm BH}}c^2$. Here, $\eta$ is assumed to be
10\%. The AGN are included in all simulations investigated here
(mergers as well as secularly evolving galaxies).

The sinks of radiation are dust grains in the ISM.  The emission from
stars and AGN undergo scattering, absorption and remission as they
propagate through the ISM.  Because the ISM substructure is not
resolved by our hydrodynamic simulations, we employ subgrid modeling
techniques which we explore in the following subsection. Eight model
cameras are set up isotropically around the galaxy to observe the
merger. The emergent flux is determined by the number of photons that
escape the galaxy in a given camera's direction.

\sunrise \ calculates the dust temperature of the grains
self-consistently, and assumes that the dust and radiation field are
in radiative equilibrium utilising the methodology of \citet{juv05}
for calculating the converged radiation field. In this methodology,
when photons are absorbed in a grid cell, the dust temperature is
updated. A new photon packet is then emitted from this cell which has
an SED equal to the difference between the SED associated with the
updated dust temperature, and that from the old one.  This procedure
is iterated upon until the radiation field has converged
\citep{jon10}. Consequently, both the AGN and the starburst are
involved in determining the dust temperature across all the grid
cells.

PAHs are emitted by the HII regions, PDRs, and the diffuse dust in the
galaxy. Carbonaceous dust grains with sizes $a < 100 $\AA \ are
assumed to have characteristics of PAH molecules. A fraction of these
PAH molecules \citep[following ][we take this value to be 50\%]{jon10}
emit a template spectrum, and the remaining fraction emit
thermally. The PAH template is a linear combination of Lorentzian
profiles, and the fraction of PAHs emitting the template is chosen to
fit {\it Spitzer} IRS observations of local galaxies
\citep{dop05,gro08,jon10}.  Their absorption is calculated utilising
the \citet{li01} cross sections.  Varying this fraction changes the
PAH luminosity (which can affect the 24 \micron \ flux density at
\z=2). We have performed test simulations exploring the range of PAH
fractions, from 0 to 1. Generally, the trends presented in this paper
are insensitive to the inclusion of PAHs, though the normalisation of
some trends can be. That is to say, the magnitude of the \tfr \ ratio
is dependent on the fraction of emitting PAHs, though the evolutionary
trends in the \tfr \ ratio are not.  Tests investigating the limits
(e.g. either 0\% or 100\% of carbonaceous grains with sizes $a < 100
$\AA \ emit as PAHs) show that the observed 24 \micron \ flux density
of our model galaxies varies at most by a factor of $\sim$2 (when
observed at \z=2.0). We note that PAHs are not explicitly destroyed by
the AGN.

The evolving dust mass is calculated assuming a constant dust to metals ratio
of 0.4, consistent with models and observations of both local Universe
galaxies, as well as those at higher redshifts
\citep{dwe98,vla98,cal08}.  The distribution and mass of interstellar
dust is determined by the galaxy evolution simulations. We use the
\citet{wei01} dust grain model with $R \equiv A_V/E_{B-V} = 3.15$, as
updated by \citet{dra07}. This model include silicates and
graphites. We note that silicate absorption features are not put in by
hand into this model - rather their occurrence is the result of
modeled radiative transfer effects.

All photometric quantities in this paper are convolved with the
appropriate instrument filter sensitivity function. Throughout this
paper, all flux densities and flux density ratios come from both
galaxies in the simulation box. While the simulation box is 100 kpc on
a side, the nuclei of the galaxies (where the bulk of the emission
arises from) never stray more than $\sim$50 kpc from each other. This
is roughly comparable to the {\it Spitzer} 24 \micron \ beam at
\z=2. Hence, during all the relevant stages of the interaction, both
galaxies would appear as a single (blended) source in {\it Spitzer}
24$\mu$m surveys.

\subsubsection{ISM Specification and Other Free Parameters}
\label{section:freeparameters}

\begin{figure}
\hspace{-1cm}
\includegraphics[angle=90,scale=0.4]{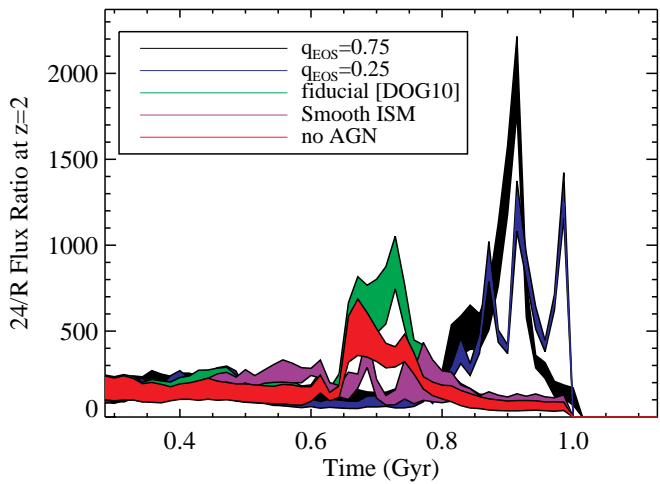}
\caption{Evolution of \tfr \ ratio for our fiducial model (DOG10)
  while varying the ISM specification and inclusion of AGN.  See text
  in \S~\ref{section:freeparameters} for
  details.\label{figure:testcase}}
\end{figure}

The ISM in our simulations is unresolved below the hydrodynamic
resolution limit of $\sim$ 100 pc.  In order to model obscuration
below these scales, we employ subgrid methods.  In the remainder of
this subsection, we discuss the effects of the subgrid assumptions on
our final results.  For the reader unfamiliar with the main results of
this paper, it may be worthwhile returning to this section after
reading through the main results to better understand the implications
of our subgrid techniques. In this section, we will refer to
Figure~\ref{figure:testcase} frequently, which summarises our
comparison of varying ISM treatments.  In this figure, we vary
assumptions regarding the ISM and AGN for our fiducial merger, DOG10
(with 80\% initial gas fraction), and plot the evolution of the \tfr
\ ratio.  The thickness of each lightcurve owes to sightline
variations in the SED.

  The young stellar clusters (with ages $< 10$ Myr) are assumed to
  reside in their nascent birthclouds, and have their spectrum
  reddened accordingly. The birthclouds are comprised of HII regions
  and photodissociation regions (PDRs), whose sizes and temperatures
  are calculated with the 1 dimensional photoionisation code \mappings
  \ \citep{gro08}.  The HII regions evolve as one-dimensional
  mass-loss bubbles \citep{cas75} and absorb much of the ionising UV
  radiation from the stellar clusters. In fact, the HII regions
  dominate the 24 \micron \ continuum (observed-frame at \z=2) flux
  density during starbursts.

The time-averaged covering fraction of PDRs is a free parameter, and
can be related to the PDR clearing time scale as $f_{\rm pdr} =
exp(-t/t_{\rm clear})$. Here, we take the PDR covering fraction to be
unity ($f_{\rm PDR}=1$) which essentially means that massive O and B
stars are assumed to be surrounded by their nascent birthclouds for
the entirety of their lives \citep[as the PDR clearing timescale is
  then longer than the lives of these stars; see Figure 6 of
][]{gro08}. While $f_{\rm PDR}$ is an unconstrained parameter, some
evidence exists that in local mergers a molecular interstellar medium
with a large volume filling factor may blanket nuclear O and B stars
for the majority of their lives \citep{dow98,sak99}. This suggests
that a large PDR covering fraction may be a reasonable assumption in
massive starbursts. The assumption of $f_{\rm PDR}=1$ is conservative:
removing the cold birthclouds typically increases the 24 \micron
\ flux density (observed at \z=2.0) (as emission from the HII regions
can escape more unhindered).

One potential issue is that the \mappings \ SEDs are parameterised in
terms of their compactness, which is dependent both on the stellar
cluster masses and the ISM pressure \citep[see e.g. equations 9-13 of
][]{gro08}.  The pressures seen in the central regions of the most
massive high-\z \ mergers exceed the largest ones available in the
\mappings \ lookup tables by a factor of $\sim$10-100.  This means
that in reality the \mappings \ HII/PDR SED will be hotter than those
at the saturated values used here \citep[see, Figure 5 of
][]{gro08}. Because the stellar/HII/PDR SEDs dominate primarily during
the peak starburst phase of a merger, the net effect of this is that
we are likely {\it underestimating} the duty cycle of the DOG phase
during merger-driven starbursts in our simulated galaxy sample.
Because our inclusion of the \mappings \ SEDs errs on the side of more
conservative models, we proceed with this caveat in mind.

Our simulations have no information about the structure of the ISM
below the resolution limit outside of a stellar cluster's birthcloud.
Effectively, what this means is that the radiative transfer models
here do not account for absorption by GMCs outside of a given stellar
cluster's birthcloud \citep{jon10}.  To place an upper limit on the
potential attenuation from the cold ISM, we have run test cases where
we assume that each grid cell is uniformly filled with dust and gas
with the mass returned from the SPH simulations.  This effectively
assumes that all dust that would be associated with GMCs in the ISM
has a volume filling fraction of unity.  Because the mass in the
birthclouds is 'borrowed' from this gas mass budget, in this special
scenario we require that $f_{\rm PDR}=0$ in order to avoid potential
double-counting issues.  In this extreme case, the ISM is optically
thick at mid-infrared (rest-frame) wavelengths, and the SED has a
larger proportion of its flux at longer wavelengths.  The purple curve
in Figure~\ref{figure:testcase} shows the evolution of the relatively
depressed \tfr \ ratio for this ``Smooth ISM'' case. We note that
while the \tfr \ ratio is in this Smooth ISM model is not as large as
the other cases, the galaxy's SED is still quite red.  In this case, a
larger fraction of the power in the SED is shifted into the FIR; as
such, this galaxy would likely be selectable by e.g. {\it Herschel}
surveys.


Our choice of equation of state may have an effect on the star
formation history and structure of the ISM in the model galaxies as
well. Nominally, we utilise the full \citet{spr03a} multi-phase EOS
which simulates the pressurisation of the ISM owing to supernovae
heating.  This is somewhat stiffer than an isothermal EOS \citep[the
  stiffest EOS is shown by the solid curve in Figure 4 of ][the isothermal
  EOS is the dashed line]{spr05a}. Referring to this figure, the
effective EOS employed can be quantified by a 'softening parameter'
such that $q_{\rm EOS} = 1$ represents the stiffest EOS which aims to
account for SNe pressurisation of the ISM.  Similarly, $q_{\rm EOS} =
0$ represents an isothermal EOS.

Because we use the stiff EOS ($q_{\rm EOS} = 1$), the gas is
relatively pressurised which retards fragmentation.  This affects both
the star formation history as well as the structure of the ISM
(galaxies with a softer EOS will be more clumpy which may affect
optical depths).  To investigate the consequences of our choice of
EOS, we have examined the evolution of a fiducial galaxy merger, DOG10
with softer EOSs ($q_{\rm EOS}=0.25$ and $q_{\rm EOS}=0.75$, linear
interpolations between isothermal and $q_{\rm EOS} = 1$).

In Figure~\ref{figure:testcase}, we show the \tfr \ lightcurves for
our fiducial model, with the aforementioned modifications to soften
the EOS (black and blue curves).  The general effect of the softening
of the EOS is to cause the ISM to become more unstable to
fragmentation, and reduce optical depths toward the central regions.
Because of this, the observed SED is hotter for the $q_{\rm EOS} =
0.75$ and $0.25$ cases than the fiducial DOG10, and the peak 24
\micron \ flux density a factor of a few higher.  Softening the
equation of state increases the likelihood that a given simulation
will be selectable as a DOG.

There is some evidence that discs at high redshift may indeed be
clumpy.  Both observations
\citep[e.g. ][]{elm07,dad08,gen06,elm09a,for09} and models
\citep[e.g.][]{bou08,dek09,elm09c,cer10,elm10} suggest that disc
galaxies at high-\z \ contain relatively large clumps of gas.
However, despite this, we choose to utilise the simulations with the
stiffest equation of state ($q_{\rm EOS}=1$).  The reasons follow.
First, with respect to simulating the formation of DOGs, this is the
middle-of-the-road choice amongst our ISM assumptions.  For example,
examination of Figure~\ref{figure:testcase} shows that while a softer
EOS more easily produces DOGs, a smooth ISM supresses them.  The model
with birthclouds (and no other cold GMCs) and a stiff EOS has a test
lightcurve which falls in between the two extremes.  Second, we can
constrain the models somewhat by the gas fractions at the time of
merger.  If we assume that mergers represent a reasonable model for
the formation of high-redshift Submillimetre Galaxies \citep[e.g. ][
  though see Dav\'e et al. 2009 for
  counter-arguments]{tac06,tac08,nar09b,nar10a}, then the gas
fractions in the model galaxies at the time of merger/peak submm flux
density should be comparable to those inferred from observations.
Indeed, \citet{nar10a} showed that the gas fractions at peak
coalescence were indeed $\sim$ 40\%, comparable to the measurements of
\citet{bou07} and \citet{tac08}.  However, because our simulations do
not include the infall of primordial gas from the IGM, there is no way
to replenish the gas once it is consumed by star formation.  Thus, by
utilising a softer EOS and increasing the level of disc
fragmentation/star formation, the models quickly run out of gas and
render the galaxy relatively gas poor during the peak submm phase.
Tests show that the simulation with $q_{\rm EOS} = 0.75(0.25)$ have
gas fractions of $\sim$20(10)\% during their final
coalescence/peak-submm phases.  In this sense, our usage of a stiff
EOS is a requirement to match certain physical properties of the most
extreme high-\z \ ULIRGs, necessitated by limitations of the models.
Moreover, in the interest of making a more direct comparison with our
previous attempts at modeling high-\z \ ULIRGs \citep[to wit, the SMG
  population ][]{nar09b,nar10a}, we utilise the stiffest EOS and
birthcloud model to remain consistent with the initial conditions
chosen for those models.

To summarise our exploration of the effects of ISM substructure on
the radiative transfer, we expect clumpiness of the ISM to exist on
three scales: on the scale of birthclouds around young stellar
clusters; in other GMCs outside of the birthclouds; and on $\sim$kpc
scales as seen in observed discs at high-\z
\ \citep[e.g. ][]{elm07,elm09a}.  Our inclusion of the \mappings
\ calculations accounts for some obscuration on the birthcloud scale,
and our parameter search with the EOS has tested the possible effects
of larger clumps on the radiative transfer. With respect to
obscuration by GMCs outside of the parent birthcloud, the current
generation of simulations only allow us to test the limiting cases of
a smooth obscuring ISM with a volume filling fraction of unity.  The
results of these tests on the models are summarised in
Figure~\ref{figure:testcase}.

Finally, we can explore the role AGN play in our model for DOG
formation.  AGN heat the surrounding dust and contribute to rest frame
8\micron \ thermal emission.  This both increases the observed \tfr
\ ratio, as well as the duty cycle over which the galaxy has extremely
red colors.  The effect of not including AGN can be seen in the red curve in
Figure~\ref{figure:testcase}.


\section{Formation and Evolution of DOGs}
\label{section:dogformation}

\subsection{Physical Evolution of a Merger}
\label{section:physical_evolution}

\begin{figure*}
\includegraphics{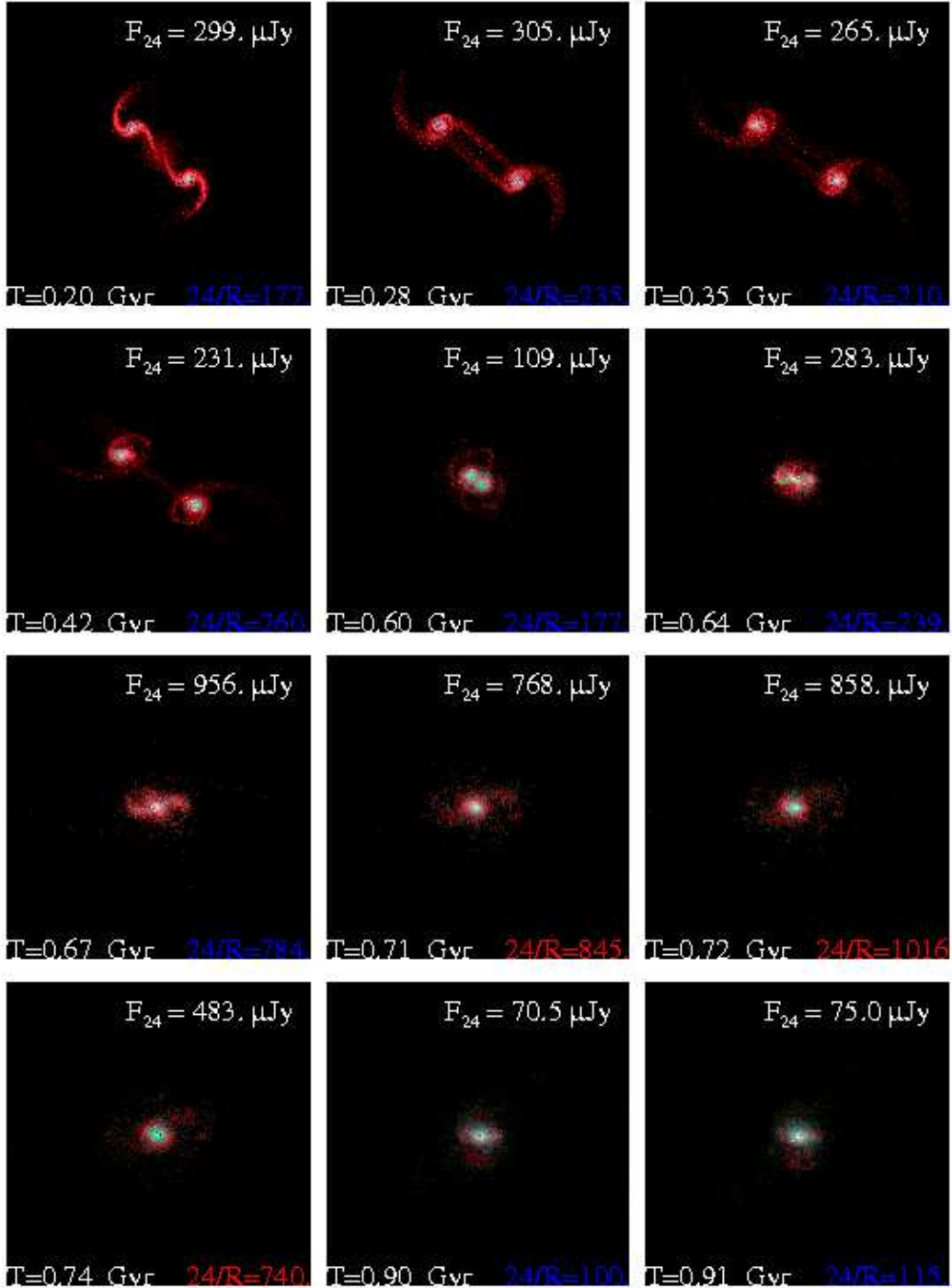}
\caption{Morphology of fiducial merger, DOG10, modeled at redshift
  \z=2. Colors are (observed-frame) MIPS 24 \micron, $R$-band, and
  $B$-band mapped onto red, green and blue.  Each panel is 100 kpc/h
  on a side, and the time stamp, 24 \micron \ flux density, and \tfr
  \ ratio is denoted in each panel. The color of the \tfr \ label
  represents if the mid-IR SED exhibits a bump or power-law nature (to
  be discussed in \S~\ref{section:bump2pl}), where blue labels show
  bump galaxies, and red labels PL galaxies.  Mergers naturally form
  DOGs. As the merger progresses, the galaxy becomes redder owing to
  increased 24 \micron \ flux density powered by the dust-enshrouded
  starburst. Upon final coalescence, when the SFR peaks, the galaxy is
  red enough that it may be observed as a DOG (\tfr $>$
  1000). Subsequent contribution from a growing AGN will allow the
  galaxy to transition from a bump DOG to a PL
  DOG.\label{figure:morph}}
\end{figure*}

The formation of ULIRGs through mergers has a long history
\citep[e.g., ][]{san88,san96}. We therefore begin our investigation of
DOGs with the ansatz that DOGs originate in gas rich galaxy mergers at
high redshift. We will then go on to explore under what circumstances
DOGs may be represented by non-merging galaxies 
(\S~\ref{section:physicalnature}), and show that as observations probe
sufficiently lower 24 \micron \ flux density limits, less extreme
conditions are sufficient to form DOGs.

We first briefly orient the reader as to the generic physical
evolution of a high-redshift binary gas rich galaxy merger. Here, we
focus on fiducial model DOG10 (Table~\ref{table:ICs}).  DOG10 is a 1:1
$M_{\rm DM} \approx 6 \times 10^{12}$ \msun merger.
Figure~\ref{figure:morph} shows the evolution of the morphology of
DOG10 as viewed from a single \sunrise \ camera.  The boxes are 100
kpc/h on a side, and the galaxy is observed at \z=2 (with the observed
24 \micron, $R$ and $B$-band emission mapped onto red, green and
blue).  DOG10 is a very similar model to as the fiducial SMG (SMG10)
studied in \citet{nar09b}. We will present a more detailed comparison
between DOGs and SMGs in \S~\ref{section:smg}.

\begin{figure}
\hspace{-.25cm}
\includegraphics[scale=0.5]{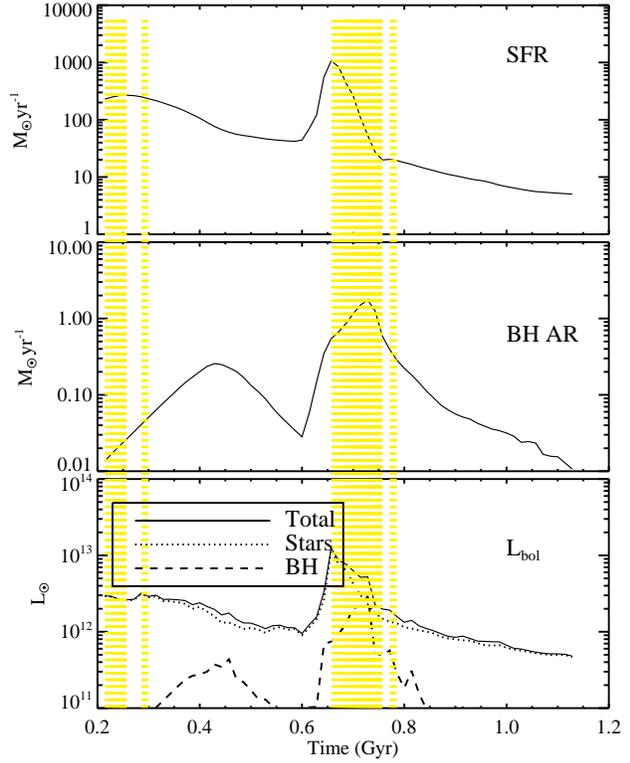}
\caption{Evolution of star formation rate, black hole accretion rate,
  and bolometric luminosity for fiducial merger model DOG10. The
  bolometric luminosity includes contributions from both stars and
  black holes. The yellow shaded region shows when the galaxy would be
  selected in a \stf $> 300 $ \microjy \ flux limited survey (the gap
  in the yellow region denotes a short time this particular simulated
  merger falls slightly below the \stf $> 300 $\microjy \ flux density
  limit). The SFR, black hole accretion and luminosity all peak near
  when the galaxy is undergoing final coalescence.  It is during this
  time that the galaxy system is expected to peak in infrared
  luminosity. We note that the black hole accretion rate and SFR are
  averages over 10 Myr intervals.
  \label{figure:bhar_sfr_lbol}}
\end{figure}

In Figure~\ref{figure:bhar_sfr_lbol}, we show the evolution of the
SFR, black hole accretion rate, and bolometric luminosity of DOG10.
The first passage of the galaxies induces a $\sim$200 \msunyr
starburst. This elevated star formation rate is largely sustained for
$\sim$5$\times$10$^8$yr while the galaxies inspiral toward final
coalescence. As such, by the time of coalescence, the galaxy system
builds up a $\sim$10$^{11}$ \msun bulge \citep{nar10a}.

When the galaxies approach for final coalescence, tidal torquing on
the gas funnels large quantities into the nucleus of the merged system
\citep{bar91,bar96,mih94a,mih96,spr05a}, inducing a massive $\sim$1000
\msunyr starburst. Concomitantly, inflows fuel central black hole
accretion. In the mergers modeled here, the black hole accretion rate
can approach $\sim$1-2 \msunyrend. The associated AGN feedback
contributes (along with gas consumption by star formation) to
terminating the starburst, and may render the black hole optically
luminous for a short period of time as feedback and gas depletion
clear sightlines \citep{spr05a,hop05b,hop05a},
accounting for the observed population of bright
quasars \citep{hop06,hop08a}. The contribution of
newly formed stars and an embedded AGN drive the $\sim$10$^{13}$\lsun
bolometric luminosity during the merger. It is during the final
merger, when the starburst, black hole accretion rate and bolometric
luminosity are all near their peak ($T \approx$ 0.65-0.75 Gyr; i.e.,
the shaded region in Figure~\ref{figure:bhar_sfr_lbol}) that we expect
the infrared luminosity of the model galaxy to peak
\citep{jon06b,you09,nar10a}.

\subsection{The Evolution of the \tfr \ Ratio in  Mergers}
\label{section:general}

\begin{figure}
\hspace{-1cm}
\includegraphics[scale=0.4,angle=90]{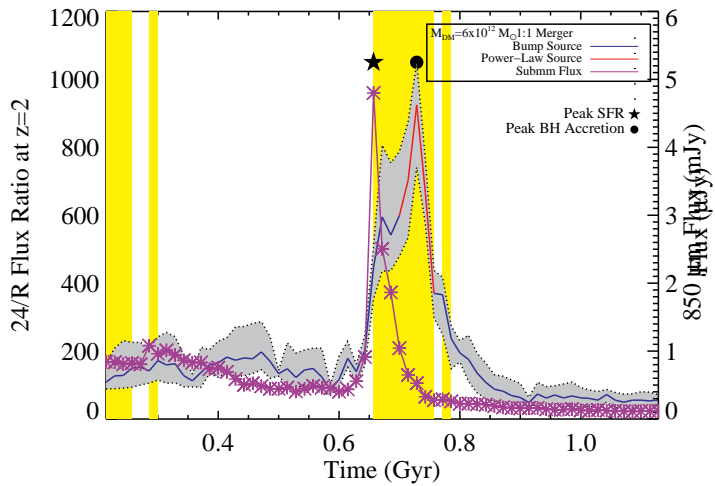}
\caption{\tfr \ and 850 \micron \ light curves for fiducial model
  DOG10, modeled at \z=2. The blue/red line is the sightline-averaged
  \tfr \ ratio as a function of time, and the color in this line
  denotes when the galaxy may be visible as a bump galaxy (blue) or PL
  galaxy (red). (See \S~\ref{section:bump2pl} for a discussion
  regarding the transition from bump to PL mid-IR SED.) The grey
  shaded region highlights the range of \tfr \ ratios seen over 8
  isotropic viewing angles, and thus reflects the sightline dependence
  of the \tfr \ ratio. The yellow shaded region shows when the
  galaxy's 24 \micron \ flux density is above a fiducial limit of 300
  \microjy.  The break in the yellow shaded region is a short phase
  when the galaxy drops just below 300 \microjy.  The black filled
  star and circle near the top show when the star formation rate and
  black hole accretion rates peak, respectively.  The 850 \micron
  \ flux density is overlaid as the purple curve, and has its units on
  the right axis. As the galaxy passes through its starburst dominated
  phase, it may be visible as both a Submillimetre Galaxy (with \sef
  $>$ 5 mJy), as well as a bump galaxy. The galaxy eventually
  transitions to becoming a PL galaxy, and may be either starburst or
  AGN dominated during this time.  \label{figure:b5e_lightcurve.smg}}
\end{figure}


\begin{figure}
\hspace{-1cm}
\includegraphics[angle=90,scale=0.4]{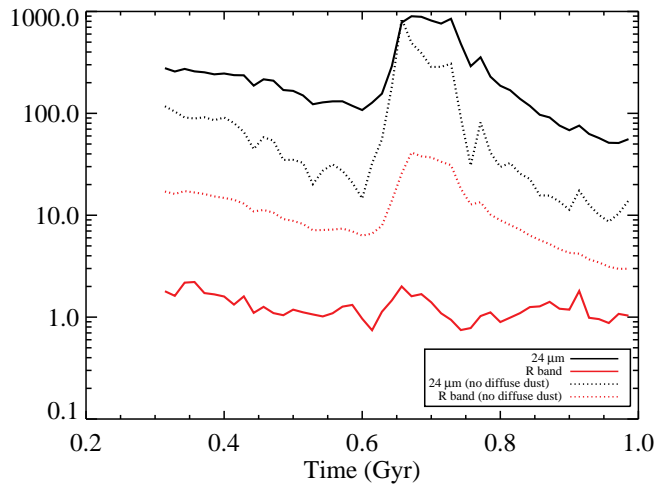}
\caption{24 \micron \ and $R$-band lightcurves for the fiducial model
  (DOG10) both including diffuse dust (solid lines) and neglecting it
  (dotted lines). We remind the reader that all models in this paper
  include diffuse dust with the exception of the dotted lightcurves in
  this figure. The models without diffuse dust still include HII
  regions surrounding stellar clusters. It is clear that the HII
  regions are responsible for a large fraction of the 24 \micron
  \ flux.  The diffuse dust is important in absorbing the UV radiation
  which escapes the HII regions; without this attenuation, the excess
  radiation which escapes the galaxy causes it to appear optically
  bright, and thus not selectable as a DOG.\label{figure:24Rphot}}
\end{figure}

 In Figure~\ref{figure:b5e_lightcurve.smg}, we
 plot the evolution of the \tfr \ ratio for fiducial merger DOG10
 placed at \z=2.  As a reminder, Figures~\ref{figure:morph} and
 ~\ref{figure:bhar_sfr_lbol} serve as a reference for the morphology,
 black hole accretion rate, SFR, and bolometric luminosity for this
 \ galaxy. The grey shaded region shows the range of \tfr \ ratios as
 seen from 8 isotropically placed cameras outside the model galaxy,
 and thus reflects the dependence of the \tfr \ ratio on viewing
 angle. The yellow shaded region shows when the galaxy could be
 detected above the nominal \stf$>$300 \microjy \ detection threshold
 at \z=2. We plot the 850 \micron \ lightcurve as well (with units on
 the right ordinate), though defer discussion of this aspect of the
 figure until \S~\ref{section:smg}). We additionally note when the
 galaxy would be a bump galaxy, and when it would be a PL galaxy (as
 the red/blue colors in the \tfr \ lightcurve), but defer discussion
 of this aspect of the galaxy's evolution to \S~\ref{section:bump2pl}.

Massive mergers at high redshift naturally produce DOGs during their
final coalescence. The galaxy appears the reddest during the
starburst/AGN activity associated with final coalescence (in our
fiducial DOG, we note that it is formally above the \tfr $> 960 $
threshold for only a brief period of time. More massive models
[e.g. DOG1] form DOGs for longer duty cycles.  DOG10 is, to some
degree, the rough minimum galaxy mass necessary in our simulations to
form a DOG).  The observed 24 \micron \ thermal emission comes from
two sources.  During the inspiral phase, warm dust in the diffuse ISM
dominates the 24 \micron \ emission. At times of elevated SFR (e.g.,
during final coalescence; $T \approx$ 0.65 Gyr), HII regions
additionally contribute a substantial fraction of the observed 24
\micron \ thermal emission allowing the galaxy to be selected in 24 \micron
\ surveys. As gas consumption and the wind from the AGN terminate both
the starburst \citep{spr05c} and black hole growth \citep{you08a}, the
\tfr \ ratio fades. Soon thereafter, the 24 \micron \ flux density
drops (with the bolometric luminosity of the galaxy) such that the
galaxy would no longer be detectable with current instruments.

 The \tfr $>1000$ DOG criterion selects star-forming galaxies which
 have large quantities of obscuring dust in the diffuse ISM.  The
 copious rest-frame UV photons produced by the starburst are
 reprocessed into rest-frame 8 \micron \ thermal radiation by dust
 both within the HII regions and in the diffuse ISM.  Though this
 point may seem trivial, it is to be underscored that $z\approx 2$
 star-forming galaxies without large columns of dust in the diffuse
 ISM may still be selected as 24 \micron-bright galaxies (owing to the
 reprocessing of some of their UV photons in HII regions), but will
 not be reddened enough in their observed \tfr\ to be selected as
 DOGs.  We demonstrate this explicitly in Figure~\ref{figure:24Rphot}, where
 we show the 24 \micron \ and $R$-band lightcurves for our fiducial
 model (DOG10) both including diffuse dust (the fiducial model) and
 neglecting it.  For the model neglecting diffuse dust, we still
 include HII regions surrounding stellar clusters. It is
 clear that the HII regions are responsible for a substantial fraction
 of the observed 24 \micron \ flux density. However, the diffuse dust
 is important in absorbing the UV radiation which escapes the HII
 regions; without this attenuation, the excess radiation which escapes
 the galaxy causes the normalisation of the \tfr \ ratio to fall
 significantly below the DOG selection criteria.


\begin{figure*}
\includegraphics[angle=90,scale=0.8]{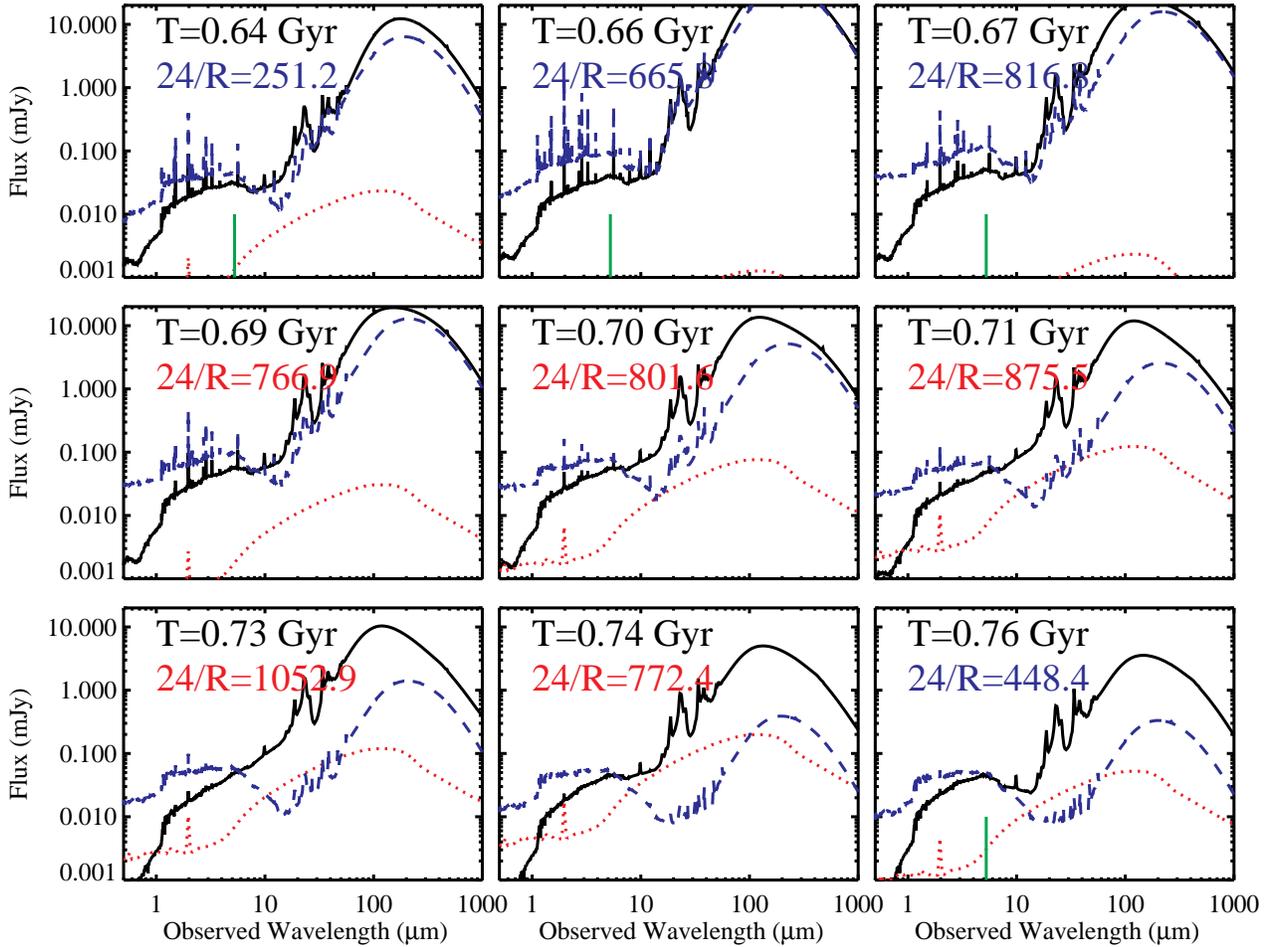}
\caption{Model SEDs for 9 snapshots of fiducial galaxy DOG10 as it
  evolves through its final coalescence DOG phase. The black line is
  the sightline-averaged observed SED (at \z=2). The blue dashed line
  represents the stellar SED (after passing through HII regions and
  PDRs), and the red dotted line is the unattenuated AGN spectrum.
  The SEDs are all redshifted to \z=2, and the abscissa is the
  observed-frame wavelength. The \tfr \ ratio is listed along with the
  time for each snapshot, and the color of the \tfr \ ratio represents
  when the galaxy would be selected as a bump galaxy (blue) versus a
  PL galaxy (red). A green vertical line is placed at the rough
  location of the redshifted stellar bump in the snapshots which are
  bump galaxy.  We note that we label the galaxy as a bump or PL
  galaxy regardless of whether it classifies as a DOG by the fiducial
  \tfr \ criteria.  The galaxy's DOG phase begins as a bump DOG (e.g.,
  $T = 0.67$ Gyr). As the merger progresses, the galaxy will be
  selected as a PL DOG, either owing to dust attenuation of the
  stellar bump (even at times when the AGN contribution is not
  particularly strong; e.g., $T = 0.70 $ Gyr), or to direct
  contribution from the AGN continuum (e.g., $T = 0.74$ Gyr). It is
  important to note that even some sources that may appear as bump
  galaxies by eye (e.g. $T=0.69 $ Gyr) may not be classified as such
  when convolving with IRAC filters. Finally, we note that while only
  a single snapshot in this Figure is formally a DOG, the duty cycle
  increases with increasing galaxy mass.  The fiducial DOG represents
  a rough lower limit to the galaxy mass which forms
  DOGs.\label{figure:six_sed}}
\end{figure*}

\subsection{Evolution of Bump DOG to Power-law DOG}
\label{section:bump2pl}

As discussed in \S~\ref{section:introduction} (see also D08), the mid-IR spectra of
DOGs come in two flavors: those characterised by a `bump' at $\sim$5
\micron, and those exhibiting a power-law rise in the SED
\citep{dey08}. Bump features in the mid-IR SED of \zsim 2 galaxies are
thought to originate from a combination of thermal emission from lower
mass stars at rest-frame 1.6 \micron, and emission from massive stars
which show a local minimum in their atmospheric opacity at this
wavelength \citep{joh88,sim99,far08}. Similarly, the PL shape in the mid-IR
spectra of \zsim 2 galaxies is canonically associated with the
dominant presence of AGN continuum (though, as we will show, the PL
SED can originate in starburst dominated galaxies).

In a merger-driven model for DOG formation, there is a natural
evolution in the mid-IR SEDs of \zsim 2 DOGs from bump-like to PL-like
as the merger progresses. For the purposes of our analysis, we fix our
model galaxies at \z=2, and define a bump galaxy utilising a simple
peak-finding method\footnote{We note that this definition of ``bump''
  versus ``powerlaw'' mid-IR SEDs is similar to that used by
  \citet{far08}, and is simpler to implement than the power-law
  fitting method used by D08.}  such that $S_{\rm 3} < S_{\rm 5}$ and
$S_{\rm 5} > S_{\rm 8}$ (where $S_{\rm 3}$, $S_{\rm 5}$, and $S_{\rm
  8}$ correspond to the filter-convolved fluxes from IRAC channels 1,2
and 4, respectively). Similarly, we define a PL galaxy as one where
$S_{\rm 3} < S_{\rm 5} < S_{\rm 8}$. Returning to
Figure~\ref{figure:b5e_lightcurve.smg}, we now see that when we
utilise this criteria for classifying the bump and PL phase of the
galaxy's evolution, there is a transition during the merger from a
bump galaxy to a PL galaxy (where the bump phase is denoted by the
blue segment of the line, and the PL phase by the red one).
Comparing with the morphology (Figure~\ref{figure:morph}), it is
evident that this transition occurs during final coalescence.

We can understand the details of the origin of the bump to PL
transition more explicitly by examining the evolution of the SEDs of
DOGs during the final coalescence stage. In
Figure~\ref{figure:six_sed}, we present the sightline-averaged SED
from fiducial galaxy DOG10 as it evolves through its final coalescence
phase. The observed SED is shown by the black solid line in each
panel. The blue dashed SED represents the input \starburst
\ SED\footnote{More specifically, the blue dashed line is the
  \starburst \ input SED after it has passed through the HII regions
  and PDRs surrounding stellar cluster utilising \mappings.}, and the
red dotted curve the input AGN template SED. Both the stellar and AGN
SEDs are unreddened by diffuse dust in the galaxy (though the stars
are reddened by HII regions and PDRs). We list the time stamp of each panel for
comparison with the morphology (Figure~\ref{figure:morph}), and global
\tfr \ light curve (Figure~\ref{figure:b5e_lightcurve.smg}), and quote
the \tfr \ ratio in each panel. The \tfr \ ratio is color coded based
on if the DOG is a bump galaxy (blue), or PL galaxy (red).

In the early stages of the DOG phase (near the beginning of final
coalescence, e.g., $T=0.67$ Gyr), the mid-IR SED shows the
characteristic stellar bump associated with a star-formation dominated
infrared luminosity. As the merger evolves, the mid-IR SED becomes
best characterised by a power-law shape. PL galaxies can arise from either
starburst-dominated or AGN-dominated DOGs. Starburst-dominated PL galaxies
owe their origin to large columns of dust obscuration associated with
the final merging. To see this, consider panel $T=0.70$ Gyr in
Figure~\ref{figure:six_sed}.  During the merger, the intrinsic stellar
flux density at $\lambda \la$ 5 \micron \ can be extinguished by up to
an order of magnitude. This can force the SED to look power-law-like,
even while the AGN contribution at these wavelengths is small. The
galaxy then evolves into an AGN-dominated PL galaxy. Here, the AGN SED
contributes substantially to the observed SED at $\lambda \ga$ 8
\micron, again giving a power-law mid-IR shape (e.g., $T=0.74$
Gyr). As is clear, then, PL mid-IR SEDs from \zsim 2 galaxies may
represent either starburst dominated or AGN dominated galaxies. The
relationship between the AGN contribution to PL mid-IR SEDs in \zsim 2
24 \micron \ sources will be explored in more detail in a forthcoming
paper (D. Narayanan et al., in prep.).

Finally, we note that in the post-merger stage, the galaxy exhibits
bluer colors with a bump-like SED.  These ULIRGs are reminiscent of
those detected which are less extreme than DOGs in deep {\it Spitzer}
surveys \citet{yan05,yan07,lon09,yan10}.

\section{The Physical Form of DOGs: Are Mergers Necessary?}
\label{section:physicalnature}

\begin{figure}
\hspace{-1cm}
\includegraphics[scale=0.4,angle=90]{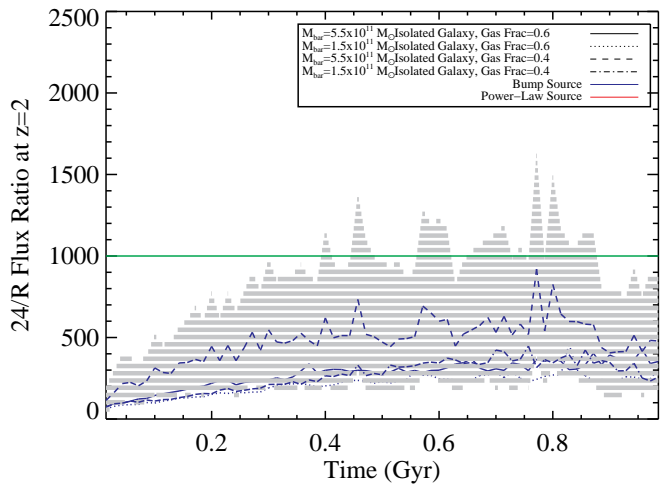}
\caption{Evolution of sightline averaged \tfr \ lightcurves for the
  most massive two of the model isolated galaxies in
  Table~\ref{table:ICs} (iDOG1 and iDOG2) are shown as blue lines.
  Both gas fractions are plotted for iDOG1 and iDOG2. The range of
  colors owing to viewing angle of the reddest galaxy is shown by the
  grey-hatched region.  While isolated galaxies typically are not red
  enough to satisfy the DOG criteria, the most massive, metal-rich
  ones certainly can when in edge-on viewing
  angles.\label{figure:isolightcurve}}

\end{figure}

Thus far, we have seen that gas rich galaxy mergers at high redshift
provide a plausible means for forming luminous DOGs. Furthermore, this
mechanism for DOG formation suggests a picture for connecting bump
DOGs and PL DOGs.  This leads to a natural question: are mergers
a necessary prerequisite for the formation of DOGs?

We begin with Figure~\ref{figure:isolightcurve}, where we show the
evolution of the \tfr \ ratio for the most massive isolated galaxies
in Table~\ref{table:ICs}.  We plot the lightcurves for both gas
initial fractions modeled - 40\% and 60\%.  The lightcurves are
sightline averaged, and we show the sightline-averaged disperion for
the reddest galaxy through the grey-shaded region.  Generally,
isolated galaxies are not quite red enough to satisfy the fiducial
\tfr $> 1000$ criteria for DOG selection.  This said, the most
massive, dust-rich ones can represent DOGs.  For example, the 40\% gas
fraction, $M_{\rm bar}\approx 5\times10^{11}$ \msun model (which
starts with nearly solar metallicity in the closed-box model)
approaches DOG-like redness.  That said, the most massive, dust-rich
isolated galaxies can be selected as DOGs when in an edge-on
configuration.

\begin{figure}
\hspace{-1cm}
\includegraphics[scale=0.4,angle=90]{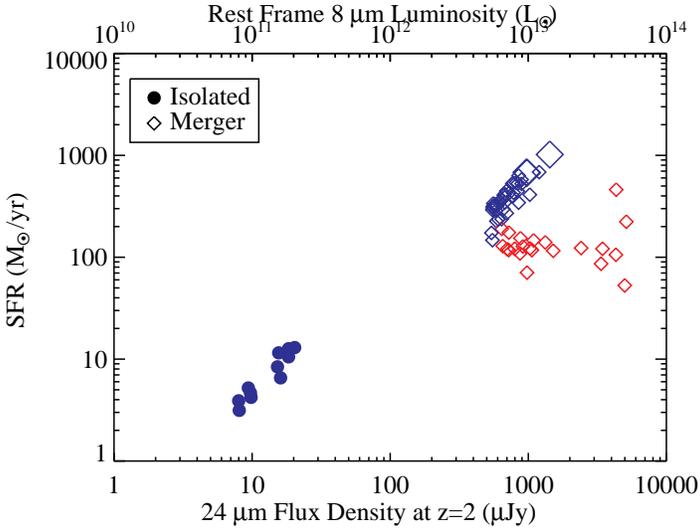}
\caption{SFR versus 24 \micron \ flux density for all snapshots of all
  model galaxies which qualify as DOGs (\tfr $>$ 1000, but with no 24
  \micron \ flux density limit). On the top axis, we show the
  rest-frame 8 \micron \ luminosity. The different symbols denote
  whether the galaxies are mergers or isolated galaxies.  The symbols
  which are double sized qualify both as DOGs, as well as \sef $>5
  $mJy, though we defer discussion of the SMG-DOG overlap until
  \S~\ref{section:smg}. There is a clear relation between the 24
  \micron \ flux density and SFR for the bump DOGs, though the
  contribution of AGN can muddy this for PL galaxies.  Generally, the
  most luminous 24 \micron \ sources require merging activity, and
  have PL mid-IR SEDs. As galaxies become less bright, they tend
  toward bump (star formation-dominated) galaxies, and are well
  described by non-mergers as well. The reader should note that the
  gap in between the mergers and isolated galaxies is artificial, and
  betrays the limited parameter space investigated in these models.
  This gap represents the transition from isolated galaxies dominating
  the bulk of the DOG population to mergers.  A broader range of
  models run at more varied masses, merger orbits and gas fractions
  would fill this space in.  It is moreover important to note that the
  flux densities on the abscissa are subject not only to the
  uncertainties present in this model (\S~\ref{section:methods}), but
  also the assumed redshift of \z=2.0. As such, the normalisation of
  this trend should be taken with the appropriate caution.
  \label{figure:sfr_24}}
\end{figure}

We can further investigate the relative role isolated discs and
mergers play toward \zsim 2 ULIRG formation in
Figure~\ref{figure:sfr_24}, where we plot the SFR versus the 24
\micron \ flux density (and rest-frame 8 \micron \ luminosity) at \z=2
for every snapshot in our simulation sample that has a \tfr \ ratio
$>$ 1000. We discriminate both between mergers and isolated galaxies,
as well as whether the galaxy is a bump or PL DOG. The larger symbols
are galaxies which would also be detectable as SMGs, though we defer
discussion of this aspect until \S~\ref{section:smg}. Mergers populate
the bulk of the luminous (\stf $>$ 300 \microjy) DOG population in
this diagram. These systems, which are extremely gas/dust rich, become
red enough to be selected as DOGs during close passages and final
coalescence (e.g., Figure~\ref{figure:b5e_lightcurve.smg}). Mergers
provide an efficient means for generating the starburst and/or AGN
activity to heat the dust sufficiently that the galaxy is visible at
24 \micron \ above 300 \microjy.  Nominally, we find a minimum SFR
$\ga 500$ \msunyr necessary when AGN activity is negligible to power
\stf $> 300$ \microjy \ sources. Similarly, for DOGs powered by AGN,
we find minimum AGN accretion rates of $\dot{M}_{\rm BH} \ga 0.5$
\msunyr \ are required to power the brightest DOGs.

At lower 24 \micron \ flux density limits, however, mergers are not
always required for the formation of DOGs. Massive discs can be red
enough to satisfy the requisite infrared-optical color ratio for DOG
selection, though are less bolometrically luminous.


 The most luminous DOGs in Figure~\ref{figure:sfr_24} tend to all be
 PL sources, consistent with observations \citep{dey08}. The isolated
 galaxies exhibit lower 24 \micron \ flux densities, and are always
 bump DOGs. At lower flux densities, the SFR is correlated with the
 observed 24 \micron \ flux density (C. Hayward et al., in prep.).  As
 the contribution of AGN becomes non negligible, though, this relation
 breaks down (e.g., for most luminous sources). While the trends in
 Figure~\ref{figure:sfr_24} are robust, we caution against
 interpreting the absolute values of the 24 \micron \ flux density too
 literally. Rest frame mid-IR PAH features contribute to the
 observed \stf \ at \zsim 2. Modulating the assumed fraction of
 carbonaceous grains which emit the PAH template can shift the
 observed 24 \micron \ flux density of a given galaxy. Changing the
 PAH fraction has the strongest relative effect on the lowest flux
 density galaxies, and at its limits (e.g., a PAH fraction of 0 or 1),
 can shift the observed flux density by up to a factor of $\sim$50\%
 (though the shift is less for more luminous sources).

To summarise, our models suggest that the most luminous DOGs (i.e.,
those with \stf $\ga 300\mu$Jy at $z\approx 2$) likely result from
mergers.  In contrast, the less luminous DOGs can result from a wider
range of (less extreme) conditions, perhaps even from secular
evolution.  Within this framework, there are a few salient points
regarding the physical nature of DOGs that appear to be general to all
DOGs selected at all flux density limits in our models.

First, in the absence of any merging activity, a DOG will never go
through a PL phase. {\it Some} merging activity is necessary to
produce the dust obscuration of UV radiation and/or AGN activity which
drive the PL shape to the observed mid-IR spectra. Mergers are
required to funnel sufficient gas to the central regions to suppress
(or overwhelm) the stellar bump (and/or fuel the AGN). As such, PL
DOGs constitute a relatively narrow class of galaxies, whereas bump
DOGs are significantly more diverse as they are well represented by
both mergers and non-mergers. This is seen explicitly in
Figure~\ref{figure:sfr_24}.

Second, while it is not shown, there is a rough minimum galaxy mass
necessary for forming DOGs. Nominally, galaxies with halos $M_{\rm
  bar} << 5\times 10^{11}$\msun will not have the requisite gas supply
to generate the starburst/AGN activity that drives the observed \tfr
\ ratio.  When mapping from baryonic masses to halo masses
(e.g. Table~\ref{table:ICs}), the halo masses are also consistent with
the clustering measurements published by \citet{bro08}. Moreover, the
halo masses are consistent with the inferred halo masses of
high-redshift galaxies of a comparable luminosity - that is, SMGs
\citep{bla04}, and quasars \citep{cro05,lid06,hop07c,she07}.

\section{Relationship of DOGs to Other High-Redshift Galaxies}
\label{section:othergalaxies}
\subsection{Submillimetre Galaxies}
\label{section:smg}

A number of clues suggest an intimate link between high-redshift 850
\micron-selected SMGs, and the DOG population \citep{dey09}.  First,
circumstantial evidence indicates physical similarities between SMGs
and DOGs. SMGs are thought to form in massive starbursts which both
observational and theoretical evidence imply may be merger-driven
\citep[e.g.,] [though see \citet{dav10} for an alternative
  interpretation]{cha03b,tac06,tac08,bau05,swi08,nar09b,nar10a}. Our
model for DOG formation suggests that luminous DOGs, as well, are
likely formed at the final stages of a merger
(\S~\ref{section:physicalnature}). Similarly, the inferred SFRs in
SMGs and PL DOGs appear to both be extremely large \citep[$\sim
  700-1500$
  \msunyrend;][]{swi04,kov06,men07,val07,dey08,bus09b,des09}. DOGs and
SMGs cluster in similar mass halos \citep{bro08}, and tentative
evidence suggests that may they contain somewhat similar mass stellar
bulges \citep[e.g. ][]{swi04,bor05,bus09b,mic09}. Second,
\citet{nar10a}'s merger-driven model for the formation of SMGs
utilised similar merger models employed in this study.  This overlap
in modes of galaxy formation (a merger-based mode for both SMG
and DOG formation) suggests not only a physical connection between
SMGs and DOGs, but possibly an evolutionary one. Finally, and perhaps
most convincingly, an analysis of the GOODS-N field by \citet{pop08b}
has found a bona fide overlap between the SMG and DOG populations:
Namely, some $\sim$30\% of SMGs have \tfr \ colors consistent with DOG
selection.

To assess the relationship between DOGs and SMGs, we now highlight the
purple 850 \micron \ flux density lightcurve overplotted in
Figure~\ref{figure:b5e_lightcurve.smg}. As the galaxies merge ($T
\approx $0.65 Gyr), radiation from the $\sim$1000-2000 \msunyr
starburst is intercepted by the cold ISM, and
re-emitted by cold dust as submillimetre-wave emission
\citep{gro08,jon10,nar10a}. At the same time, diffuse dust in the
galaxy heavily reddens the optical light, rendering the galaxy visible
as a DOG. The model DOG10 would be classified as a bump galaxy at the
exact same time as it would be visible as an SMG
(Figure~\ref{figure:b5e_lightcurve.smg}). The formal DOG phase follows
after the SMG phase in this model, though we note that numerous models
in Table~\ref{table:ICs} show an overlap of the two phases.  After the
SMG phase, the fiducial galaxy transitions from a bump DOG to a PL
DOG. Because the SMG phase and DOG phase of the galaxy are both
triggered by the merger, the phenomena are intimately related.

\begin{figure}
\hspace{-1cm}
\includegraphics[scale=0.4,angle=90]{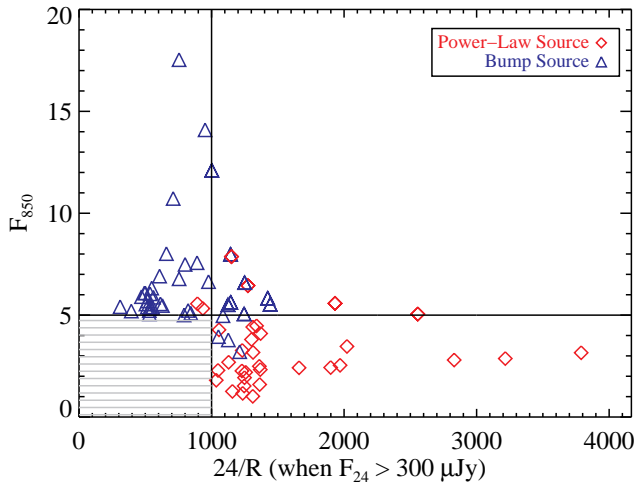}
\caption{850 \micron \ flux density versus \tfr \ ratio for all 24
  \micron-selected galaxies in our model sample. Blue triangles show
  when the galaxies would be selected as bump DOGs, and red diamonds
  as PL DOGs.  Only galaxies which are either SMGs or DOGs are
  plotted.  The vertical and horizontal lines denote the DOG \tfr
  \ selection criteria and the SMG \sef \ selection criteria. While
  there is significant overlap between the SMG (\sef $>$ 5 mJy) and
  DOG population, numerous DOGs are not detectable in the
  submillimetre, and vice-versa. It is important to recognise that the
  region of space where \tfr $<$ 1000 \ and \sef $<$ 5 mJy (shown by
  the grey hatched region) is not populated by construction, as in
  this space galaxies are neither DOGs nor SMGs. In reality, a large
  number of galaxies reside in this portion of \sef-\tfr
  \ space. \label{figure:S850_DOG}}
\end{figure}

 While Figure~\ref{figure:b5e_lightcurve.smg} is instructive as to a
 general sequence of events, it should not be construed to mean that
 all SMGs would be selected as DOGs, much less bump DOGs. Merger
 orbit, viewing angle, and mass can shift the exact timing and
 magnitude of events to make the overlap less straightforward. To show
 this, in Figure~\ref{figure:S850_DOG}, we plot the 850 \micron \ flux
 density from every snapshot in every model in Table~\ref{table:ICs},
 making two selection cuts: the galaxy has to either be classified as
 an SMG (\sef$>$5 mJy), or a DOG (\tfr $>$ 1000). We additionally note
 when the galaxies would be visible as bump sources, or PL DOGs. There
 is a significant overlap between the galaxy populations. A few
 features are apparent.

First, while there is overlap, many SMGs will not be selected as DOGs,
and vice versa. Second, SMGs may be either bump or PL sources. The
fact that some SMGs have bump mid-IR SEDs makes sense as the 850
\micron \ flux which SMGs are selected on is powered by absorption of
stellar light by cold clouds \citep{nar10a}, and are generally thought
to have their bolometric luminosity powered by star formation
\citep{cha04,ale05a,val07,pop08,you08b,men09}. That SMGs can also be
PL DOGs may seem counterintuitive given that the dominant power source
in SMGs is thought to be star formation. Recall, however, that some PL
sources derive their mid-IR SED shapes from extremely heavy dust
extinction at 5 \micron, even while the AGN continuum does not
dominate the mid-IR flux density. While the PL DOG phase is roughly
contemporaneous with the rise in AGN continuum, they are not always
exactly coincident in time.

SMGs will typically overlap with the more (24 \micron) luminous
DOGs. This owes to the fact that in our model, SMGs derive from
massive mergers \citep[see Figure 2 from ][]{nar10a}. The overlap of SMGs with
high 24 \micron \ flux density DOGs is apparent from
Figure~\ref{figure:sfr_24}, where we highlight the symbols with double
size which are galaxies which are selectable both as DOGs, as well as
\sef $> 5 $ mJy DOGs.

Finally, we caution that Figure~\ref{figure:S850_DOG} should not be
interpreted as a measure of the relative fractions of SMGs which are
DOGs or vice versa. To construct those statistics will require a
convolution of the SMG and DOG duty cycles with cosmological galaxy
merger rates, a task deferred to a future study.

\subsection{\bzk \  Galaxies}

The \bzk \ selection criteria have been successfully used to find both
star-forming and passively evolving galaxies at redshifts $1.4\la z
\la 2.5$ \citep[e.g., ][]{dad04}. These samples overlap the DOG and
SMG redshift ranges, and therefore provide an interesting comparison.

\begin{figure}
\hspace{-1cm} 
\includegraphics[scale=0.4,angle=90]{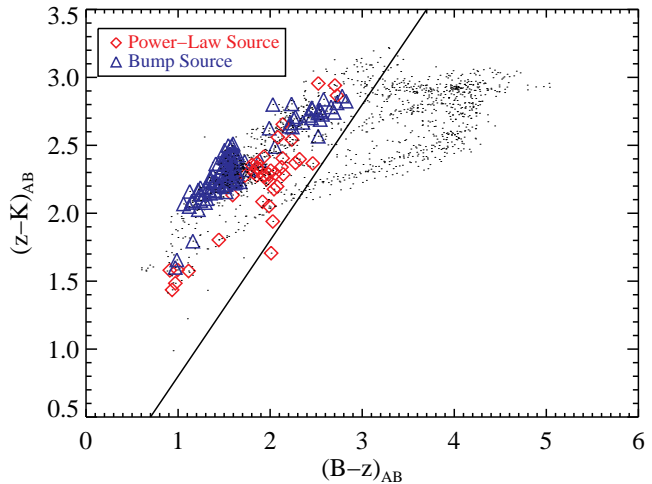}
\caption{DOGs on \bzk \ color-color plot. Bump and PL DOGs are
  denoted, and the selection criteria for star-forming galaxies
  ($(z-K)_{\rm AB}-(B-z)_{\rm AB} > -0.2$) is shown as a solid line
  (star-forming \bzk \ galaxies lie above the line). We note the
  location of all galaxies in our simulation sample (whether or not
  they are DOGs) in $(z-K)-(B-z)$ space by dots. For the purposes of
  comparisons with observational data sets, DOGs in this plot are
  selected with \stf $>$ 100 \microjy. Most model DOGs are selectable
  as \bzk \ star-forming galaxies, consistent with studies of the
  overlap between GOODS-N \bzk \ galaxies and DOGs
  \citep{pop08b}.  \label{figure:bzk}}
\end{figure}

In Figure~\ref{figure:bzk}, we plot all DOGs in our simulation sample
on the \bzk \ color-color plot. In order to compare with
\citet{pop08b}'s analysis of GOODS-N DOGs (selected with \stf $>$ 100
\microjy) in \bzk \ space, we include all DOGs with \stf $>$ 100
\microjy \ at \z = 2. We separate bump and PL DOGs, and overplot the
rough criteria for star-forming galaxies ($(z-K)_{\rm AB}-(B-z)_{\rm
  AB} > -0.2$) devised by \citet{dad04} (such that galaxies which lie
above the line are star-forming \bzk \ galaxies). Finally, we note the
location of all galaxies in our simulation sample (whether or not they
are DOGs) in $(z-K) - (B-z)$ space by dots.

Most DOGs in our sample are selected as \bzk \ galaxies. This result
is not surprising. \citet{dad05} find that the majority of \bzk
\ galaxies in the GOODS-N field are 24 \micron \ detected.  In more
direct comparison to these model results, \citet{pop08b} find that
nearly all DOGs in the GOODS-N field are selected as BzK galaxies. The
results here are in good agreement with the dataset of \citet{pop08b},
spanning a similar range in $(B-z)$ and $(z-K)$ colors.

There are two important caveats. First, while most DOGs can be
selected as \bzk \ galaxies, this does not mean that all \bzk
\ galaxies will be selected as DOGs. Observationally, \citet{pop08b}
find that only 12\% of \bzk \ galaxies are selectable as DOGs. Second,
even though most DOGs fall into the star-forming region of the \bzk
\ color-color plot, they may not be detectable at optical
wavelengths. Because the \bzk \ criteria covers a broad range of
star-forming galaxies, the DOG selection criteria selects a dusty
subset of the \bzk's.

\section{Observational Consequences of the Model}
\label{section:observations}

In order to assess the validity of our models, we compare the
simulated and observed SEDs.  We then make testable predictions of
these models, and provide analysis tools via public SED
templates. When comparing to observations, we focus primarily on the
dynamic range of fluxes and color ratios. Matching the observed
distributions (e.g., luminosity functions, color distributions)
requires cosmological galaxy formation simulations, and will be
presented in forthcoming work.

\subsection{Comparisons to Observations: Optical-mm Wave SED}
\label{section:sed}
\begin{figure*}
\includegraphics[angle=90,scale=0.75]{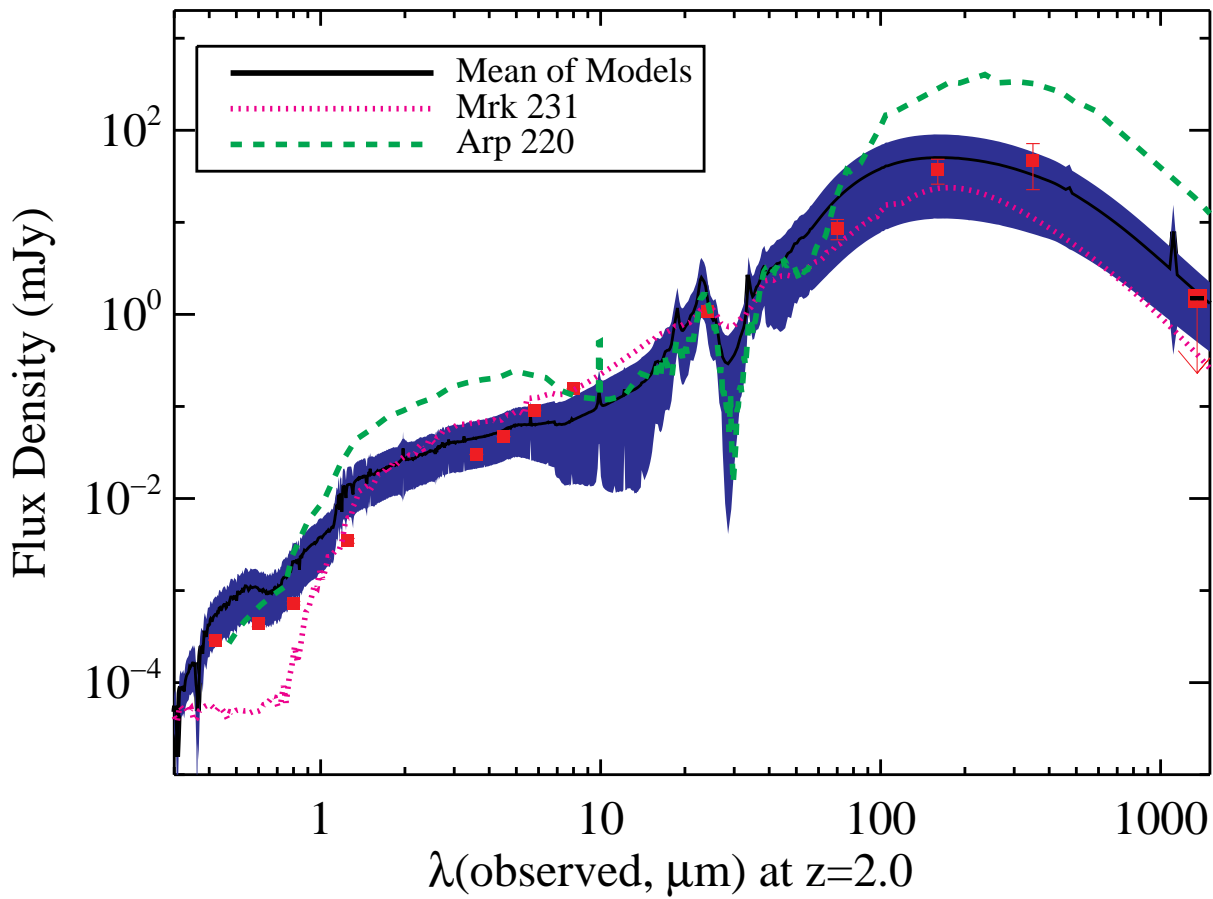}
\caption{Simulated SEDs redshifted to \z=2 for model galaxies. The SED
  was calculated for all snapshots that satisfied the DOG selection
  criteria (\tfr $>$ 1000, \stf $>$ 300 \microjy) for four model
  galaxies which sample the full mass and merger mass ratio range of
  our simulation sample (DOGs 1, 4, 7 and 10): The 1$\sigma$
  dispersion amongst these snapshots is shown by the blue shaded
  region, and the mean by the black solid lines. The model SED is
  compared to the compiled observational data points (red) by Bussmann
  et al. (in prep) for all \stf $> 300 $\microjy \ DOGs with
  spectroscopic redshifts $1.75<z<2.25$.  If error bars are not shown,
  it is because they are smaller than the symbol representing the data
  point. In order to best compare with the Bussmann et al. sample, we
  did a Monte Carlo sampling of our model SEDs that enforced that we
  had 50\% bump DOGs and 50\% PL DOGs in the sample, comparable to the
  relative breakdown of mid-IR SED types in the Bussmann et
  al. sample. The magenta dotted lines and green dashed lines are
  template SEDs for local galaxies Mrk 231 and Arp 220, scaled in flux
  density to match the observed Bussmann et al. 24 \micron \ data
  point, and scaled to \z=2.0.  The modeled SED and observed data
  generally correspond well, suggesting a reasonable match between the
  modeled stellar mass and flux and obscuring dust columns and
  temperature in the modeled galaxies and DOGs in nature. That said,
  the modeled mid-IR SED shows some disagreement with the observed
  IRAC points (especially at 8 \micron). Neither the Mrk 231 and Arp
  220 templates provide good matches to the observed
  data. \label{figure:dog_sed}}
\end{figure*}

Observed SEDs of DOGs with spectroscopic redshifts have been compiled
by Bussmann et al. (2009, in prep).  In Figure~\ref{figure:dog_sed},
we present the model SEDs for four merger-driven DOGs which span the
(mass) range of galaxies which form \stf $> 300 $\microjy \ DOGs
(DOG1, DOG4, DOG7 and DOG10) and sample the full mass ratios of
simulated mergers. We note that this is a simple mass (and mass ratio)
scaling in simulated mergers, but aside from this, the initialisation
of each merger (including galaxy orbit) remains identical. The SED is
modeled for every snapshot while the galaxy is a DOG (above \stf $>
300 $ \microjy), and is averaged over all camera angles. The black
line represents the time and camera-averaged SED, and the blue shaded
region denotes the 1$\sigma$ temporal dispersion for each model. The
SED is modeled to be at \z=2.0. The red squares show the mean of all
data in the Bussmann et al. sample for DOGs above \stf $> 300
$\microjy \ within the redshift range $1.75 < z < 2.25$, and the error
bars denote the propagated errors. In order to more accurately compare
with the Bussmann et al. sample, we performed a Monte Carlo drawing of
the galaxies in our simulation sample that forced our final sample to
be comprised of 50\% bump DOGs and 50\% PL DOGs in the composite
SED. This is comparable to the breakdown of mid-IR SED types in the
observational dataset of Bussmann et al. We note, however, that simply
including all DOG snapshots in our simulation sample makes little
difference on the final composite SED. We additionally include
template SEDs for local galaxies Mrk 231 \citep[][ R. Chary, private
  communication]{arm07,cha01}, and Arp 220 \citep{rie09}, each scaled
to match the Bussmann et al. MIPS 24 \micron \ point and redshifted to
\z=2.0.

Generally, there is a reasonable match between the observed and
modeled SEDs. The correspondence between the simulated optical, NIR
and FIR SEDs, and those observed in DOGs implies that the models
presented here have a similar UV flux from newly formed stars, stellar
mass, and dust obscuration column as DOGs in nature. There are
relatively minor discrepancies in the observed-frame NIR and mid-IR.
The Mrk 231 and Arp 220 templates both have weaknesses in
matching the observed data. The Arp 220 template overpredicts both the
rest-frame optical-NIR and FIR flux densities. Similarly, the Mrk 231
SED underpredicts the rest-frame UV substantially. Given the
relatively improved match between these models and the observed data
points (as opposed to the local templates), we provide the model SEDs
as templates in \S~\ref{section:templates}.

\begin{figure}
\hspace{-1cm}
\includegraphics[angle=90,scale=0.4]{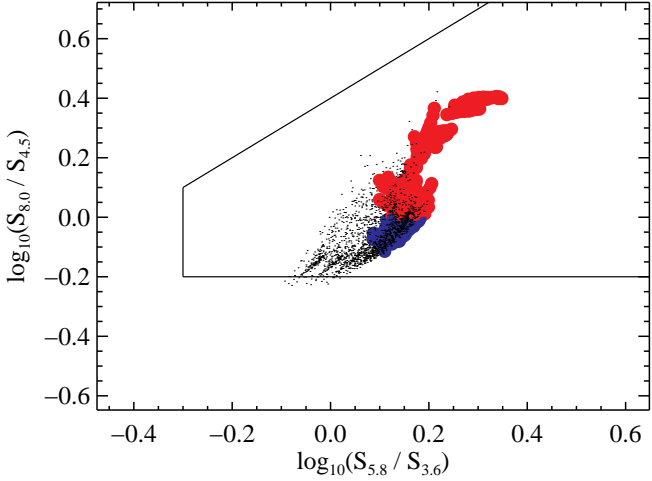}
\caption{Mid-IR color-color plot for all model galaxies which would be
  selected above a 24 \micron $>$ 300 \microjy \ flux density
  cut. DOGs are shown as colored points, with bump DOGs as blue
  symbols and PL DOGs as red symbols. Black dots represent 24 \micron
  \ selected galaxies which do not qualify as DOGs. The model galaxies
  are incrementally redshifted from \z=1.5-2.4, and plotted when they
  are selectable as DOGs.  The solid lines enclose the typical region
  of color-color space occupied by AGN-dominated galaxies as found by
  \citet{lac04}. The locus of PL DOGs lie in the top-right region of
  these plots \citep[e.g., ][]{lac04} owing to their rising mid-IR
  SEDs.  The model DOGs lie roughly in the same region of the mid-IR
  as observed DOGs \citep{des09}, though span a smaller dynamic
  range. \label{figure:irac_colorcolor}}
\end{figure}

Figure~\ref{figure:irac_colorcolor} shows where the simulated galaxies
fall on an IRAC color-color diagram.  Here, we have taken our
simulated galaxies, and redshifted them between \z=1.5 to 2.4
incrementally (in steps of $\Delta z$=0.1), and plotted all galaxies
with a \tfr \ ratio $> 1000$, and \stf \ $> 300 $\microjy.  The dots
show the galaxies which have \stf \ $> 300 $\microjy, but do not have
the requisite \tfr \ ratio to be selected as a DOG. The PL DOGs show
redder colors than the bump DOGs, which in turn are redder (in
5.8/3.6) than the non-DOG 24\micron\ sources. The simulated colors of
all models lie roughly in the same region as the observations
\citep{des09,lac04}, but span a smaller dynamic range. The solid lines
enclose the region noted by \citet{lac04} to typically contain
AGN. Both observed DOGs \citep{des09} and these models tend to fall
primarily in the \citet{lac04} AGN-wedge. While indeed DOGs powered by
AGN fall into the AGN-wedge, many starburst-dominated galaxies appear
to as well.

\subsection{Testable Predictions}
Our modeling suggests that the bright DOGs (\stf $> 300 $\microjy \ at
\z=2) are the result of mergers and represent a phase of massive
galaxy evolution following a sub-mm bright phase. If true, the models
can be used to predict observational tests of this scenario. We note
that these tests do not apply to the fainter DOGs, which  are a
more diverse population.

\subsubsection{The Physical Masses of DOGs and Their Relation to SMGs}

\begin{figure}
\hspace{-1cm}
\includegraphics[angle=90,scale=0.4]{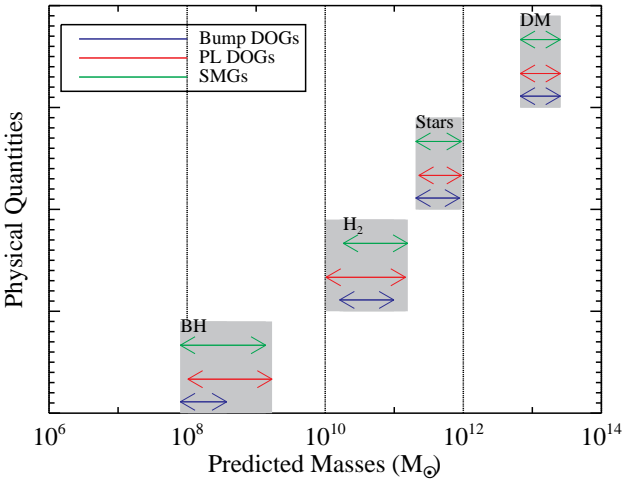}
\caption{Predicted range of physical properties of DOGs and SMGs (\stf
  $> 5$ mJy). Shown are black hole masses, \htwo \ masses, stellar
  masses and halo masses for galaxies which would be selected as DOGs
  between redshifts \z=1.6 and 2.4. Blue arrows denote bump DOGs and
  red arrows, PL DOGs.  The vertical lines are simply to guide the eye
  to the axes. DOGs are relatively massive galaxies, with PL DOGs
  tending toward the higher mass end of the mass distribution. It is
  important to note that owing to relative abundances, most observed
  DOGs are likely to occupy the lower end of the predicted
  ranges. Because similar types of galaxy (mergers) represent SMGs and
  bright DOGs, they share an overlap in physical properties.  Because
  only halo masses are known for DOGs, the remaining physical
  properties serve as a strong testable prediction of these
  models. \label{figure:dog_physical_prediction}}
\end{figure}

 In Figure~\ref{figure:dog_physical_prediction}, we plot the range of
 black hole, stellar, dark matter, and \htwo \ masses for all galaxies
 in our simulated sample at times they may be a DOG, discriminating
 both bump and PL DOGs (\stf $>$ 300 \microjy).  This prediction is
 made for galaxies that would be selected as DOGs between a redshift
 range of \z=1.6-2.4. We additionally show the same prediction for all
 galaxies in our simulations which would be selected as SMGs (with
 \sef $>$ 5 mJy).  The black hole, stellar, and halo masses are
 extracted directly from the hydrodynamic simulations. We note that
 the halo masses are predicted by construction as the mass of the
 simulated galaxies is set by the initial halo mass. While the SPH
 simulations do not explicitly follow the \htwo \ molecular gas, we
 infer its mass by considering all of the dense, star-forming gas to
 be neutral, and extracting the \htwo/HI fraction utilising the
 \citet{bli06} interstellar pressure-based prescription
 \citep{nar09b}.

Because we are considering a \stf $>$ 300 \microjy \ flux density
limit, all of the DOGs in this plot formed from mergers. Mergers
produce the bulk ($\sim$ 90\%) of their stellar mass prior to the
final merging \citep[][T.J. Cox et al. 2009,
  submitted]{nar08a,nar09}. Consequently, DOGs which form at the final
coalescence of a merger will typically have already formed the bulk of
their stellar mass. As such, the bump DOGs and PL DOGs in our
simulations tend to have relatively large stellar masses, $M_\star
\ga$ 10$^{11}$, comparable to observations \citep{lon09,bus09}.
Because similar types of galaxies form SMGs and bright (\stf $> 300
$\microjy) DOGs (e.g. massive mergers), there is substantial overlap
between the physical properties of bright DOGs and SMGs.  Generally,
only the halo masses are well-constrained for DOGs.  As such, the
remaining physical properties (stellar, BH, gas masses) shown in
Figure~\ref{figure:dog_physical_prediction} serve as a prediction.

The growth story for black holes is markedly different from that of
the stars. Black holes in mergers undergo the bulk of their
growth during the final merging \citep[see, e.g., Figure 5 from
][]{nar08a}. As such, because the PL phase tends to follow the bump
phase in these high-redshift merger models, on average, the black
holes in PL DOGs will be larger than those in bump DOGs.

We can utilise this growth history of stellar and black hole masses in
mergers to provide a test for our evolutionary model for mergers to
transition from an SMG phase to a PL DOG phase
(e.g., Figure~\ref{figure:b5e_lightcurve.smg}).  Merger remnants, in
our simulations, lie on the \magorrian \ relation owing to the
termination of the starburst event by AGN feedback, and
self-regulation of black hole growth
\citep{dim05,rob06b,hop07b,you08a,li07}.  Because the black holes grow
substantially during the final coalescence of a merger, one might
expect that the black holes in SMGs will be smaller, on average than
those in PL DOGs if SMGs are to evolve into PL DOGs as these models
suggest (Figure~\ref{figure:b5e_lightcurve.smg}). In contrast, their
stellar masses will be roughly comparable as the stellar masses are
already in place at final merging.

\begin{figure}
\hspace{-1cm}
\includegraphics[scale=0.4,angle=90]{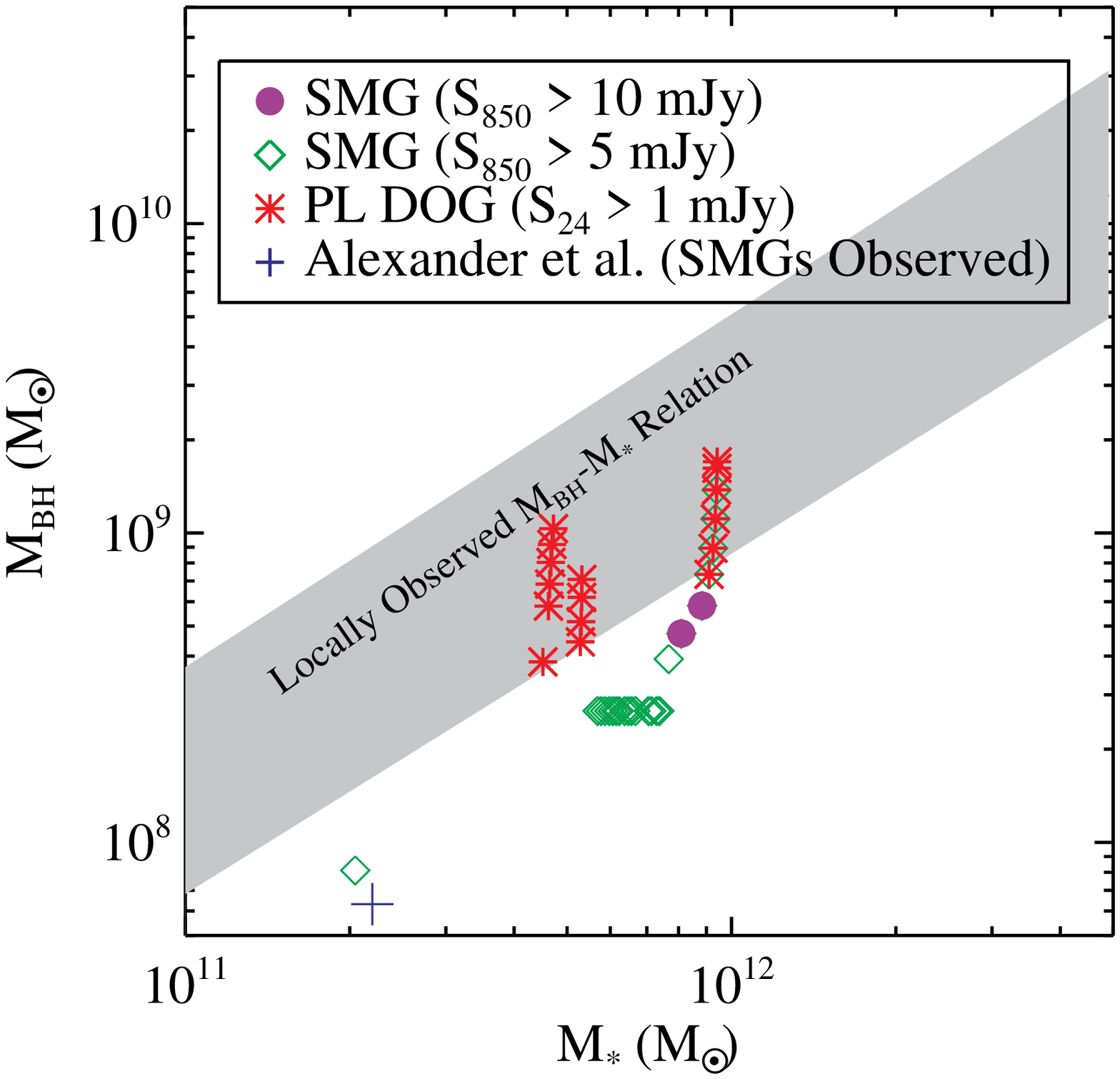}
\caption{Bright (\stf $>$ 1 mJy) PL DOGs and SMGs (in two flux density
  bins) on the present day \magorrian \ relation.  The purple circles
  denote bright SMGs, the green diamonds the average SMGs, the blue
  cross the results from observed SMGs by \citet{ale08}, the red
  asterixes bright DOGs, and the grey shaded region the locally
  observed \magorrian \ relation (including dispersion). In our model,
  SMGs typically precede the PL DOG phase in mergers. During this
  time, while the stellar mass is roughly in place, the black holes
  undergo significant growth, placing SMGs and PL DOGs at different
  locations on the \magorrian \ relation. Observations pinpointing the
  location of SMGs and PL DOGs on the \magorrian \ relation will serve
  as a test for our model connecting SMGs and DOGs.
  \label{figure:magorrian}}
\end{figure}

To test this, we compare the location of all bright SMGs and bright
(\stf $>$ 1 mJy) DOGs in our simulation sample on the \magorrian
\ relation (Figure~\ref{figure:magorrian}). We bin the SMGs into
average (\sef $> 5$ mJy) and bright (\sef $> 10$ mJy) bins. The
extremely bright (\sef $> 10 $ mJy) SMGs constitute a cleaner test as
they are always galaxies at the peak of their starburst, whereas lower
flux density sources (e.g., \sef $\approx$ 5 mJy) SMGs can be either
lower mass galaxies at the peak of their starburst, or more luminous
galaxies both pre and post-starburst
\citep{nar09b,nar10a}. Nevertheless, we include lower flux density
(\sef $> 5$ mJy) SMGs for the purposes of comparison with existing
data sets.  

Figure~\ref{figure:magorrian} shows a direct prediction for our
model. Most SMGs will lie below the present-day \magorrian \ relation
as their black holes have significant growth they will undergo
throughout the remainder of the merger \citep[e.g., ][]{ale08}. They
typically have stellar masses ranging from $10^{11} - 10^{12}$
\msunend, with the more common SMGs tending toward the lower end of
this mass range\footnote{Indeed comparison to the compilation of SMGs
  by \citet{ale08} (the blue cross) shows that the average observed
  SMG is more similar to our SMG models which form lower mass
  galaxies. }. Bright PL DOGs, which occur contemporaneously with or
later than SMGs in a merger's evolution (e.g.,
Figure~\ref{figure:b5e_lightcurve.smg}) will range from falling below
the present-day \magorrian \ relation, to directly on/above it.

\subsubsection{$T_{\rm Dust}$ of DOGs}
Forthcoming facilities which probe both sides of the peak of the SED
of DOGs (e.g., {\it Herschel}, ALMA) will place strong constraints on the
effective dust temperature. These dust temperatures will serve as a
test for the predicted SEDs of DOGs from our models.

\begin{figure*}
\hspace{-1cm}
\includegraphics[angle=90,scale=0.7]{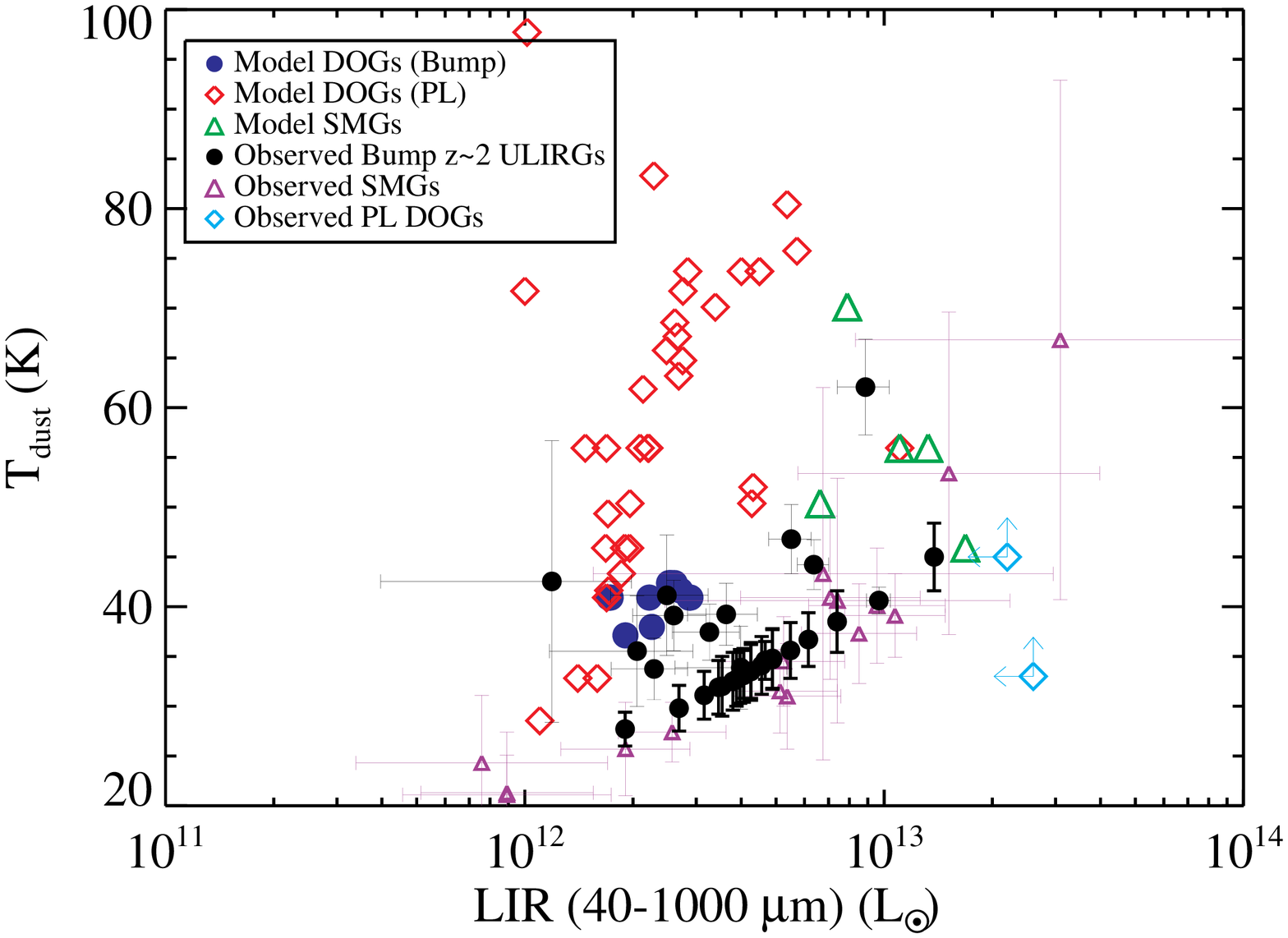}
\caption{Predicted $T_{\rm dust}$ for DOGs above \stf $>$ 100 \microjy
  \ and SMGs with \sef $> 5$ mJy as a function of \lir (40-1000
  \micron). The simulated $T_{\rm dust}$ is calculated via the peak of
  the sightline averaged SED. There is a general trend for brighter
  DOGs to exhibit hotter dust temperatures. PL DOGs have hotter dust
  temperatures than bump DOGs, as well as SMGs.  For comparison, we
  show observational data points from \citet{you09b,kov10}, \citet{kov06}
  and \citet{bus09b} which represent 1 mm-selected \zsim 2 bump
  ULIRGs, \zsim 2 SMGs, and PL DOGs (lower limits) respectively. There
  is a reasonable agreement with the high-\z observations and these
  simulations. \label{figure:tdust}}

\end{figure*}

In Figure~\ref{figure:tdust}, we plot the predicted dust temperature
of our model DOGs as a function of their FIR luminosity (40-1000
\micron). The dust temperature is derived by converting the rest-frame
wavelength of the SED peak to a temperature (using Wien's displacement
law), and is calculated for every snapshot where \tfr $>$ 1000 and
\stf $>$ 100 \microjy \ (we utilise the lower flux density limit for
completeness). For reference, we show all times when the model
galaxies would be visible as SMGs (\sef $> 5 $mJy as well). While
deriving $T_{\rm dust}$ from the full SED is a different approach than
that typically used by observers (who fit the observed data with SEDs
varying $T_{\rm dust}$ and the dust spectral index, $\beta$, which are
degenerate with one another), we prefer this method: Utilising the
entire model SED avoids any assumptions about $\beta$, which is
observed to fall within a relatively large range of $\beta = 1-2$ for
\zsim 2 SMGs and DOGs \citep{kov06,bus09b}.

For star-forming galaxies (including bump galaxies and SMGs), there is
a general trend for brighter sources to have hotter dust temperatures
(which is largely a selection effect for galaxies selected for cold
dust emission). PL DOGs typically have hotter dust temperatures than
bump DOGs and SMGs. These features are consistent with observations of
both high redshift galaxies
\citep[e.g.,][]{bus09b,cas09,you09b,kov06,kov10} and local ULIRGs
\citep[e.g.,][]{san96}.  For example, Figure~\ref{figure:tdust} also
shows \zsim 2 bump ULIRGs from \citet{you09b} and \citet{kov10}, SMGs
from \citet{kov06}, and two lower limits from PL DOGs from
\citet{bus09b} as black circles, purple triangles, and cyan diamonds,
respectively. The ULIRGs from the \citet{you09b} sample were selected
based on their stellar bump \citep{hua09}, and are thus bump galaxies.
For the bump ULIRGs of \citet{kov10}, we divided the luminosity by a
factor of 2 to convert from \lir \ to $L_{\rm 40-1000}$, based on the
approximate conversion for Mrk 231. Both the bump ULIRGs and SMGs span
a similar range in $T_{\rm dust}$-\lir \ space as these
simulations. Dust temperature determinations for bright PL DOGs will
provide a more stringent test of these models as they are expected to
show a larger range of dust temperatures than either SMGs or bump
galaxies.

\subsubsection{The CO Line Widths from DOGs}
\citet{nar08c,nar09b} found that CO traces the global
kinematics of a galaxy. The CO line widths can constrain the masses
and dynamical states of galaxies which form DOGs, and thus constitute a
testable prediction of these models. For example, while in an isolated
disc galaxy, the line width is reflective of the virial velocity of
the galaxy.  In a merger, the CO linewidths may increase by a factor
$\sqrt{2}-2$, owing to multiple galaxies being in the observational
beam and the temporary disruption of molecular discs. In this picture,
galaxies which owe their origin to mergers will exhibit larger CO
line widths than secularly evolving galaxies. 

In order to examine the predicted CO line widths from DOGs, we employ
\turtlebeach, a 3D non-local thermodynamic equilibrium molecular line
radiative transfer code \citep{nar06b,nar08a,nar09b}.  \turtlebeach
\ is an iterative Monte Carlo code which considers both radiative and
collisional processes in determining the molecular excitation (and,
hence, source functions), and assumes statistical equilibrium in the
levels \citep{nar08a,nar09b}. For galaxy-scale simulations, we
implement a subgrid procedure for including a mass spectrum of giant
molecular clouds as singular isothermal spheres, which follow the
Galactic mass spectrum \citep{ros05} and mass-radius relationship
\citep[e.g., ][]{sol87}. Once the level populations have converged
from the Monte Carlo code, the equation of radiative transfer is
integrated through the grid to build the emergent spectrum.

\begin{figure}
\hspace{-1cm}
\includegraphics[angle=90,scale=0.4]{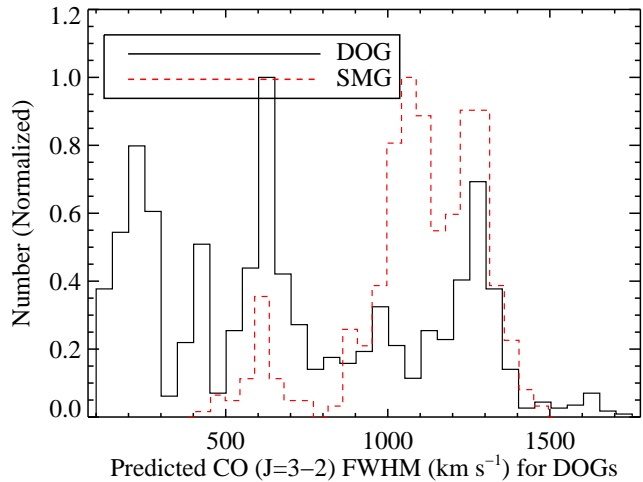}
\caption{Predicted CO (J=3-2) line widths (FWHM) for model galaxies
  when they qualify as DOGs and/or SMGs. The latter galaxy population
  is selected via a submillimetre flux density cut, \sef $>$ 5
  mJy. Line widths in galaxies are a function of their mass and
  dynamical state. Because DOGs can be either minor mergers (which may
  maintain a disc-like gas configuration), or major mergers (which are
  dynamically hot), the CO line widths span a large range (FWHM $\sim
  100-1500 $ \kmsend). In contrast, because our modeled SMGs all owe
  their origin to massive mergers, they typically have large CO line
  widths \citep{nar09b}. (See text for further details.)  It is
  important to note that the histograms are not true distributions,
  but rather simply the range of expected CO line widths from these
  galaxy populations. Because line width roughly scales with mass, the
  average galaxy will have line widths toward the lower end of the
  predicted ranges.\label{figure:dog_cofwhm}}
\end{figure}

In Figure~\ref{figure:dog_cofwhm}, we plot the range of CO (J=3-2)
FWHM for all model galaxies in Table~\ref{table:ICs} when they
classify as a luminous (\stf $>$ 300 \microjy) DOG at \z=2. As a
comparison, we plot the CO FWHM from the same model galaxies when they
would classify as an SMG (\sef $>$ 5 mJy). The line widths are viewing
angle dependent, and, as such, we include the line widths from 100
random sightlines for each model galaxy. The ranges are plotted in
histogram form, with each random sightline of each snapshot qualifying
as an individual galaxy for the histogram. It is crucial to note that
the histogram does {\it not} represent a probability distribution
function. Rather, the ordinate values are meaningless as the galaxies
are not drawn from a cosmological simulation representing true mass
functions and merger rates. The robust quantity to take from
Figure~\ref{figure:dog_cofwhm} is the {\it range} of predicted CO line
widths for DOGs and SMGs.

The line widths from DOGs show a large range, from $\sim$100 \kms to
$> 1500 $ \kmsend. This owes to the fact that both relatively minor
mergers (e.g., DOG9) as well as massive major mergers (e.g., DOG1)
form DOGs. The former are in a relatively dynamically relaxed state,
and can be viewed along some viewing angles in a face-on
configuration, causing small line widths. The latter almost always
have relatively large line widths. The CO line widths from SMGs, in
contrast, are typically quite large, ranging from $\sim$500-1500
\kmsend. This owes to their origin in massive major mergers
\citep{nar10a}.  Hence, DOGs which may be in a disc-like configuration
at times, are expected to espouse a much larger range of CO line
widths than SMGs. Tentative observational evidence suggests that this
may indeed be the case \citep[][]{gre05,tac06,tac08,yan10}.

\subsection{Analysis Tools: IR SED Templates}
\label{section:templates}

We provide public SED templates of \zsim 2 DOGs both to serve as a
testable prediction for these models, as well as an analysis tool for
forthcoming {\it JWST}, {\it Herschel}, 1.1mm and ALMA observations.  The
templates are at model spectral resolution covering 968 wavelengths
between approximately 0.3 micron and 3 mm. The SEDs are provided at
\z=2.0, the mean redshift for DOGs, and are calculated utilising every
model in the simulation sample. The templates are provided as a
function of total infrared luminosity (1-1000 \micron). The
templates as well as an IDL script to read them in (and redshift them
to arbitrary redshifts) may be found at
http://www.cfa.harvard.edu/$\sim$dnarayan/DOG\_SED\_Templates.html.

\section{Discussion}
\label{section:discussion}

\subsection{The Role of DOGs in Galaxy Evolution}

The models presented here suggest that the bright DOGs are the result
of mergers, whereas the fainter DOGs can have diverse origins.  At the
high flux density end (\stf $\ga$ 300 \microjy\ at $z\approx 2$), DOGs
represent the coalescence phase of massive, gas rich galaxy mergers
residing in large ($M_{\rm DM} \approx 10^{13}$ \msunend)
halos. Bolometrically, these galaxies are amongst the most luminous at
\zsim 2, and form stars at rates of SFR $\ga$ 500 \msunyrend. Forming
bulges with masses a few $\times 10^{11}$ \msun during the course of
their mergers, these galaxies likely are precursors to massive
present-day spheroids. These merger-driven DOGs share an intimate link
with SMGs, and represent objects actively transitioning between having
their IR luminosity dominated by reprocessed stellar radiation to
being dominated by a dust-enshrouded AGN. These massive mergers
provide a natural avenue for supermassive black hole growth, fueling
the growth of $\sim 10^9$ \msun black holes.

The bright DOGs (\stf $> 300 $\microjy) typically are undergoing
either a starburst (here, roughly $\ga$500 \msunyrend) or contain an actively
accreting AGN ($\dot{M}_{\rm BH} \ga 0.5$ \msunyrend). These phenomena
are most easily achieved via mergers. As was seen in
Figure~\ref{figure:sfr_24}, 1:1 mergers are not necessary, and more
``minor'' mergers (here, 1:3) can form bright DOGs. Given the increasing
relative number densities of minor mergers
\citep[e.g., ][]{fak08,hop09d}, it may be that these sorts of
mergers dominate bright DOG formation. It is interesting to note that
the higher mass-ratio mergers (e.g., 1:10) are not dissimilar
qualitatively from galaxies which are accreting massive amounts of
cold gas from their parent halo at \zsim 2 \citep[e.g., ``Stream Fed
  Galaxies''; ][]{ker05,bir07,dek08,dek09,ker09}.

At lower (24 \micron) flux densities (\stf $\la $100 \microjy), DOGs
can result from either galaxy mergers or galaxies undergoing secular
evolution, not in any active stage of heightened star formation or AGN
activity.  These galaxies tend to have lower SFRs ($\sim$50-100
\msunyrend), though are of comparable mass as their merging
counterparts.  In order to be selected as DOGs, these galaxies must be
viewed in an edge-on configuration which maximizes the dust
reddenning. These galaxies may have overlap with the high redshift
\bzk \ population, and (in these simulations) always have their
infrared luminosity dominated by star formation. DOGs form a
continuum, then, of galaxies, parameterised by their 24 \micron
\ luminosity.  As such, there is, of course, overlap between merging
and non-merging DOG populations in this sequence.

Quantifying the exact number statistics of types of galaxies (e.g
galaxy mass, merger mass ratio) in a given luminosity bin requires
convolving light curves with cosmological galaxy merger rates, a task
that is outside the scope of this paper. That said, these model
results do point us toward some understanding of the physical form of
DOGs at various luminosities.  Figure~\ref{figure:sfr_24} demonstrates
that mergers dominate the high flux density end, whereas isolated disc
galaxies contribute to the fainter population.  Generically, we can
say that the lower flux density DOGs, commonly identified in the deep,
narrow field surveys, are likely to be a different breed than higher
flux density DOGs which are typically uncovered by the wide-area {\it
  Spitzer} surveys.

\begin{figure}
\hspace{-1 cm}
\includegraphics[angle=90,scale=0.4]{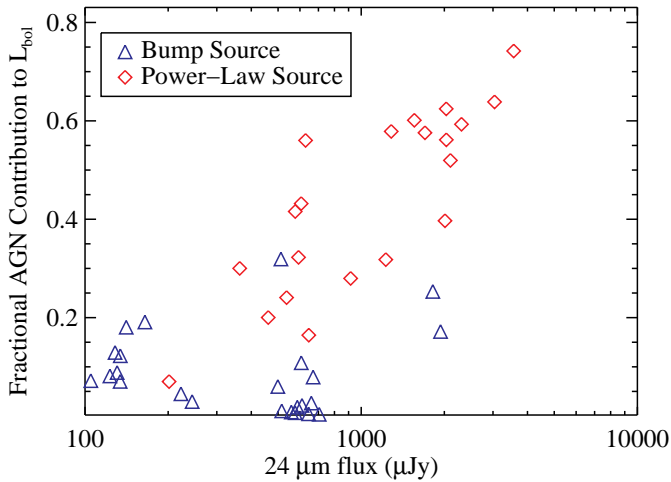}
\caption{Fractional AGN contribution to bolometric luminosity for DOGs
  (above \stf $>$ 100 \microjy) versus 24 \micron \ flux
  density. Generally, more luminous DOGs tend to have larger AGN
  contributions to their total luminosity.
  \label{figure:dog_sb_agn}}
\end{figure}

The idea that DOGs have luminosity-dependent physical properties has
already been seen in observational data sets. For example, there is
some evidence that more luminous DOGs are more likely to harbor
AGN. D08 noted that at increasing 24 \micron \ flux densities, a
larger fraction of galaxies are X-ray detected and have PL mid-IR
SEDs. Similarly, a number of studies have shown that luminous DOGs
(\stf $ \ga 500 $\microjy) may be Compton thick and contain obscured
AGN \citep{fio08,fio09,geo09,sac09}.  In our simulations, we see that
brighter sources have increasing contribution to the bolometric
luminosity by AGN (Figure~\ref{figure:dog_sb_agn}). While assessing
whether or not these AGN are Compton thick is outside the scope of
this work, we can make some crude estimates along these lines.  If we
assume that the molecular ISM in our simulations is uniformly
distributed in the nuclear regions \citep[as is thought to be the case
  for local mergers; ][]{dow98}, and not terribly clumpy, then we find
some sightlines towards the nucleus may be marginally Compton thick
($N_{\rm H} \sim 10^{24}$ cm$^{-2}$). However, a full investigation
assessing the sightline dependence of column densities as well as the
assumptions regarding the cold-phase clumping will be necessary to
fully address whether these simulated DOGs are Compton thick.

Further evidence for luminosity-dependent physical properties in DOGs
comes from \citet{mel09}, who found an increasing concentration and
decreasing half-light radius with increasing 24 \micron \ flux density
in DOGs.  Finally, \citet{bro08} found that DOGs exhibit
luminosity-dependent clustering, such that the most luminous DOGs in
their sample (\stf $> 600 $\microjy) tend to reside in $\sim 10 ^{13}$
\msun halos, while less luminous (\stf $> 300$ \microjy) DOGs are
associated with lower mass $M_{\rm DM} \approx 2 \times 10^{12}$ \msun
halos. All of these are generally consistent with our picture for more
luminous DOGs to be represented by high-redshift mergers, and less
luminous DOGs displaying less star formation and AGN activity
(Figure~\ref{figure:dog_sb_agn}).

\section{Summary and Conclusions}
\label{section:conclusions}

Utilising a combination of polychromatic radiative transfer
calculations and hydrodynamic simulations of high-redshift galaxy
evolution, we have formulated a physical model for the formation and
evolution of \zsim 2 Dust-Obscured Galaxies. We have utilised the
very similar models as those employed in previous studies aimed at
investigating the formation and evolution of a similar class of \zsim
2 ULIRGS: the Submillimetre Galaxy population.

 While the uncertainty in the models is dominated by small-scale
 physics which is unresolved by the simulations, our methodology
 allows us to make the following general conclusions regarding the
 origin of DOGs, and their relationship to other high-\z \ ULIRGs:

\begin{itemize}

\item DOGs are a diverse class of galaxies, ranging from secularly
  evolving star-forming disc galaxies (forming stars at $\sim$50-100
  \msunyrend) to extreme gas-rich galaxy mergers forming stars at
  $\ga$ 1000 \msunyrend. 

\item The most luminous DOGs (\stf $\ga$ 300 \microjy) are well
  represented by mergers of massive ($M_{\rm DM} \approx 5 \times
  10^{12}-10^{13}$ \msunend) galaxies. These galaxies are actively
  transitioning from being starburst dominated to being AGN dominated.
  At decreasing 24 \micron \ flux densities (\stf $\la$ 100 \microjy),
  DOGs may either be galaxy mergers, or secularly evolving disc
  galaxies with more modest ($\sim$50-100 \msunyrend) SFRs.

\item Luminous, merger-driven DOGs naturally transition from having a
  bump-like mid-IR SED to a power-law (PL) shape, as the dominant
  power source transitions from star formation to the central
  AGN. That said, there is an overlap period where star-formation
  dominated sources can appear as PL galaxies.

\item The most luminous DOGs assemble both significant stellar masses
  ($M_\star \approx 10^{11}$ \msunend), as well as contribute toward
  the growth of supermassive ($M_{\rm BH} \approx 10^9$ \msunend)
  black holes. 

\item Merger-driven DOGs overlap with the \zsim 2 Submillimetre Galaxy
  population. SMGs generally represent the earlier,
  starburst-dominated phase of the DOGs evolution. This can be tested
  via the location of SMGs and DOGs on the \magorrian \ relation.

\end{itemize}

In advance of {\it Herschel}, {\it JWST} and ALMA, we provide the following
testable predictions for our model of DOG formation:

\begin{itemize}

\item We quantify the expected range of black hole, \htwo, stellar,
  and halo masses for luminous (\stf $>$ 300 \microjy) DOGs as well as SMGs.

\item We provide dust temperatures for DOGs selected at various 24
  \micron \ flux density limits.

\item We detail the location of SMGs and bright (\stf $>$ 1 mJy) DOGs
  on the \magorrian \ relation as a test for the modeled evolutionary
  scenario that SMGs evolve into PL DOGs.

\item We provide a prediction for the CO line widths from DOGs, and
  suggest that they will be of order the largest observed line widths
  at high redshift, comparable to \zsim 2 SMGs.

\end{itemize}

Finally, we provide model SED templates for \z=2 DOGs as a function of
total infrared luminosity (\lir = 1-1000 \micron). These are provided
publicly, along with an IDL script for easily reading in the templates
at http://www.cfa.harvard.edu/~dnarayan/DOG\_SED\_Templates.html

\section*{Acknowledgments} We are grateful to Ranga-Ram Chary, Vandana Desai, 
Brandon Kelly, Kai Noeske, Jason Melbourne and Alex Pope for helpful
conversations.  D.N. thanks the NOAO in Tucson for hospitality, where
part of this study was conducted. The authors are grateful to the
W.M. Keck Foundation for hosting the Napa Galaxy Evolution workshop
where the ideas for much of this project came about. AD is supported
by NOAO, which is operated by the Association of Universities for
Research in Astronomy (AURA) under a cooperative agreement with the
National Science Foundation. C.C.H. was funded by an NSF Graduate
Research Fellowship. T.C. and M.B. acknowledge support from the
W.M. Keck Foundation. RSB acknowledges financial assistance from HST
grant GO10890, which is provided by NASA through a grant from the
Space Telescope Science Institute which is operated by AURA under NASA
contract NAS5-26555. PJ was supported by programs HST-AR-10678 and
10958, provided by NASA through a grant from the Space Telescope
Science Institute, which is operated by the Association of
Universities for Research in Astronomy, Incorporated, under NASA
contract NAS5-26555, and by {\it Spitzer} Theory Grant 30183 from the Jet
Propulsion Laboratory. Support for PFH was provided by the Miller
Institute for Basic Research in Science, University of California,
Berkeley.  The simulations in this paper were run on the Odyssey
cluster supported by the Harvard FAS Research Computing Group.

\bibliographystyle{mn2emod} 
\bibliography{/Users/dnarayanan/paper/refs}

\begin{thebibliography}{}

\bibitem[\protect\citeauthoryear{{Alexander}, {Bauer}, {Chapman}, {Smail},
  {Blain}, {Brandt} \& {Ivison}}{{Alexander} et~al.}{2005}]{ale05a}
{Alexander} D.~M.,  {Bauer} F.~E.,  {Chapman} S.~C.,  {Smail} I.,  {Blain}
  A.~W.,  {Brandt} W.~N., {Ivison} R.~J.,  2005, \apj, 632, 736

\bibitem[\protect\citeauthoryear{{Alexander} et~al.,}{{Alexander}
  et~al.}{2008}]{ale08}
{Alexander} D.~M. et~al., 2008, \aj, 135, 1968

\bibitem[\protect\citeauthoryear{{Armus}, {Charmandaris}, {Bernard-Salas},
  {Spoon}, {Marshall}, {Higdon}, {Desai}, {Teplitz}, {Hao}, {Devost}, {Brandl},
  {Wu}, {Sloan}, {Soifer}, {Houck} \& {Herter}}{{Armus} et~al.}{2007}]{arm07}
{Armus} L.,  {Charmandaris} V.,  {Bernard-Salas} J. et~al., 2007, \apj, 656,
  148

\bibitem[\protect\citeauthoryear{{Barnes} \& {Hernquist}}{{Barnes} \&
  {Hernquist}}{1996}]{bar96}
{Barnes} J.~E., {Hernquist} L.,  1996, \apj, 471, 115

\bibitem[\protect\citeauthoryear{{Barnes} \& {Hernquist}}{{Barnes} \&
  {Hernquist}}{1991}]{bar91}
{Barnes} J.~E., {Hernquist} L.~E.,  1991, \apjl, 370, L65

\bibitem[\protect\citeauthoryear{{Baugh} et~al.,}{{Baugh}
  et~al.}{2005}]{bau05}
{Baugh} C.~M. et~al., 2005, \mnras, 356, 1191

\bibitem[\protect\citeauthoryear{{Birnboim}, {Dekel} \& {Neistein}}{{Birnboim}
  et~al.}{2007}]{bir07}
{Birnboim} Y.,  {Dekel} A., {Neistein} E.,  2007, \mnras, 380, 339

\bibitem[\protect\citeauthoryear{{Blain}, {Chapman}, {Smail} \&
  {Ivison}}{{Blain} et~al.}{2004}]{bla04}
{Blain} A.~W.,  {Chapman} S.~C.,  {Smail} I., {Ivison} R.,  2004, \apj, 611,
  725

\bibitem[\protect\citeauthoryear{{Blitz} \& {Rosolowsky}}{{Blitz} \&
  {Rosolowsky}}{2006}]{bli06}
{Blitz} L., {Rosolowsky} E.,  2006, \apj, 650, 933

\bibitem[\protect\citeauthoryear{{Borys}, {Smail}, {Chapman}, {Blain},
  {Alexander} \& {Ivison}}{{Borys} et~al.}{2005}]{bor05}
{Borys} C.,  {Smail} I.,  {Chapman} S.~C.,  {Blain} A.~W.,  {Alexander} D.~M.,
  {Ivison} R.~J.,  2005, \apj, 635, 853

\bibitem[\protect\citeauthoryear{{Bouch{\'e}} et~al.,}{{Bouch{\'e}}
  et~al.}{2007}]{bou07}
{Bouch{\'e}} N. et~al., 2007, \apj, 671, 303

\bibitem[\protect\citeauthoryear{{Bournaud}, {Daddi}, {Elmegreen}, {Elmegreen},
  {Nesvadba}, {Vanzella}, {Di Matteo}, {Le Tiran}, {Lehnert} \&
  {Elbaz}}{{Bournaud} et~al.}{2008}]{bou08}
{Bournaud} F.,  {Daddi} E.,  {Elmegreen} B.~G. et~al., 2008, \aap, 486, 741

\bibitem[\protect\citeauthoryear{{Bouwens}, {Thompson}, {Illingworth}, {Franx},
  {van Dokkum}, {Fan}, {Dickinson}, {Eisenstein} \& {Rieke}}{{Bouwens}
  et~al.}{2004}]{bou04}
{Bouwens} R.~J.,  {Thompson} R.~I.,  {Illingworth} G.~D. et~al., 2004, \apjl,
  616, L79

\bibitem[\protect\citeauthoryear{{Brand}, {Dey}, {Weedman}, {Desai}, {Le
  Floc'h}, {Jannuzi}, {Soifer}, {Brown}, {Eisenhardt}, {Gorjian}, {Papovich},
  {Smith}, {Willner} \& {Cool}}{{Brand} et~al.}{2006}]{bra06}
{Brand} K.,  {Dey} A.,  {Weedman} D. et~al., 2006, \apj, 644, 143

\bibitem[\protect\citeauthoryear{{Brodwin}, {Dey}, {Brown}, {Pope}, {Armus},
  {Bussmann}, {Desai}, {Jannuzi} \& {Le Floc'h}}{{Brodwin}
  et~al.}{2008}]{bro08}
{Brodwin} M.,  {Dey} A.,  {Brown} M.~J.~I. et~al., 2008, \apjl, 687, L65

\bibitem[\protect\citeauthoryear{{Bullock}, {Kolatt}, {Sigad}, {Somerville},
  {Kravtsov}, {Klypin}, {Primack} \& {Dekel}}{{Bullock} et~al.}{2001}]{bul01}
{Bullock} J.~S.,  {Kolatt} T.~S.,  {Sigad} Y.,  {Somerville} R.~S.,  {Kravtsov}
  A.~V.,  {Klypin} A.~A.,  {Primack} J.~R., {Dekel} A.,  2001, \mnras, 321, 559

\bibitem[\protect\citeauthoryear{{Bussmann}, {Dey}, {Borys}, {Desai},
  {Jannuzi}, {Le Floc'h}, {Melbourne}, {Sheth} \& {Soifer}}{{Bussmann}
  et~al.}{2009}]{bus09b}
{Bussmann} R.~S.,  {Dey} A.,  {Borys} C. et~al., 2009, ArXiv e-prints/0909.2650

\bibitem[\protect\citeauthoryear{{Bussmann}, {Dey}, {Lotz}, {Armus}, {Brand},
  {Brown}, {Desai}, {Eisenhardt}, {Higdon}, {Higdon}, {Jannuzi}, {LeFloc'h},
  {Melbourne}, {Soifer} \& {Weedman}}{{Bussmann} et~al.}{2009}]{bus09}
{Bussmann} R.~S.,  {Dey} A.,  {Lotz} J. et~al., 2009, \apj, 693, 750

\bibitem[\protect\citeauthoryear{{Calura}, {Pipino} \& {Matteucci}}{{Calura}
  et~al.}{2008}]{cal08}
{Calura} F.,  {Pipino} A., {Matteucci} F.,  2008, \aap, 479, 669

\bibitem[\protect\citeauthoryear{{Caputi}, {Lagache}, {Yan}, {Dole},
  {Bavouzet}, {Le Floc'h}, {Choi}, {Helou} \& {Reddy}}{{Caputi}
  et~al.}{2007}]{cap07}
{Caputi} K.~I.,  {Lagache} G.,  {Yan} L. et~al., 2007, \apj, 660, 97

\bibitem[\protect\citeauthoryear{{Casey}, {Chapman}, {Beswick}, {Biggs},
  {Blain}, {Hainline}, {Ivison}, {Muxlow} \& {Smail}}{{Casey}
  et~al.}{2009}]{cas09}
{Casey} C.~M.,  {Chapman} S.~C.,  {Beswick} R.~J. et~al., 2009, \mnras, pp
  1231--+

\bibitem[\protect\citeauthoryear{{Castor}, {McCray} \& {Weaver}}{{Castor}
  et~al.}{1975}]{cas75}
{Castor} J.,  {McCray} R., {Weaver} R.,  1975, \apjl, 200, L107

\bibitem[\protect\citeauthoryear{{Ceverino}, {Dekel} \& {Bournaud}}{{Ceverino}
  et~al.}{2010}]{cer10}
{Ceverino} D.,  {Dekel} A., {Bournaud} F.,  2010, \mnras, pp 440--+

\bibitem[\protect\citeauthoryear{{Chapman} \& {Casey}}{{Chapman} \&
  {Casey}}{2009}]{cha09}
{Chapman} S.~C., {Casey} C.~M.,  2009, \mnras, 398, 1615

\bibitem[\protect\citeauthoryear{{Chapman} et~al.,}{{Chapman}
  et~al.}{2004}]{cha04}
{Chapman} S.~C. et~al., 2004, \apj, 611, 732

\bibitem[\protect\citeauthoryear{{Chapman}, {Windhorst}, {Odewahn}, {Yan} \&
  {Conselice}}{{Chapman} et~al.}{2003}]{cha03b}
{Chapman} S.~C.,  {Windhorst} R.,  {Odewahn} S.,  {Yan} H., {Conselice} C.,
  2003, \apj, 599, 92

\bibitem[\protect\citeauthoryear{{Chary} \& {Elbaz}}{{Chary} \&
  {Elbaz}}{2001}]{cha01}
{Chary} R., {Elbaz} D.,  2001, \apj, 556, 562

\bibitem[\protect\citeauthoryear{{Cox}, {Jonsson}, {Primack} \&
  {Somerville}}{{Cox} et~al.}{2006}]{cox06a}
{Cox} T.~J.,  {Jonsson} P.,  {Primack} J.~R., {Somerville} R.~S.,  2006,
  \mnras, 373, 1013

\bibitem[\protect\citeauthoryear{{Croom}, {Boyle}, {Shanks}, {Smith}, {Miller},
  {Outram}, {Loaring}, {Hoyle} \& {da {\^A}ngela}}{{Croom}
  et~al.}{2005}]{cro05}
{Croom} S.~M.,  {Boyle} B.~J.,  {Shanks} T. et~al., 2005, \mnras, 356, 415

\bibitem[\protect\citeauthoryear{{Daddi}, {Dannerbauer}, {Elbaz}, {Dickinson},
  {Morrison}, {Stern} \& {Ravindranath}}{{Daddi} et~al.}{2008}]{dad08}
{Daddi} E.,  {Dannerbauer} H.,  {Elbaz} D.,  {Dickinson} M.,  {Morrison} G.,
  {Stern} D., {Ravindranath} S.,  2008, \apjl, 673, L21

\bibitem[\protect\citeauthoryear{{Daddi} et~al.,}{{Daddi}
  et~al.}{2004}]{dad04}
{Daddi} E. et~al., 2004, \apj, 617, 746

\bibitem[\protect\citeauthoryear{{Daddi} et~al.,}{{Daddi}
  et~al.}{2005}]{dad05}
{Daddi} E. et~al., 2005, \apjl, 631, L13

\bibitem[\protect\citeauthoryear{{Daddi} et~al.,}{{Daddi}
  et~al.}{2009}]{dad10a}
{Daddi} E. et~al., 2009, ArXiv/0911.2776

\bibitem[\protect\citeauthoryear{{Dav{\'e}}, {Finlator}, {Oppenheimer},
  {Fardal}, {Katz}, {Kere{\v s}} \& {Weinberg}}{{Dav{\'e}}
  et~al.}{2009}]{dav10}
{Dav{\'e}} R.,  {Finlator} K.,  {Oppenheimer} B.~D.,  {Fardal} M.,  {Katz} N.,
  {Kere{\v s}} D., {Weinberg} D.~H.,  2009, MNRAS Accepted: arXiv/0909.4078

\bibitem[\protect\citeauthoryear{{Dekel}, {Birnboim}, {Engel}, {Freundlich},
  {Goerdt}, {Mumcuoglu}, {Neistein}, {Pichon}, {Teyssier} \& {Zinger}}{{Dekel}
  et~al.}{2008}]{dek08}
{Dekel} A.,  {Birnboim} Y.,  {Engel} G. et~al., 2008, ArXiv e-prints

\bibitem[\protect\citeauthoryear{{Dekel}, {Birnboim}, {Engel}, {Freundlich},
  {Goerdt}, {Mumcuoglu}, {Neistein}, {Pichon}, {Teyssier} \& {Zinger}}{{Dekel}
  et~al.}{2009}]{dek09}
{Dekel} A.,  {Birnboim} Y.,  {Engel} G. et~al., 2009, \nat, 457, 451

\bibitem[\protect\citeauthoryear{{Desai}, {Soifer}, {Dey}, {LeFloc'h}, {Armus},
  {Brand}, {Brown}, {Brodwin}, {Jannuzi}, {Houck}, {Weedman}, {Ashby},
  {Gonzalez}, {Huang}, {Smith}, {Teplitz}, {Willner} \& {Melbourne}}{{Desai}
  et~al.}{2009}]{des09}
{Desai} V.,  {Soifer} B.~T.,  {Dey} A. et~al., 2009, \apj, 700, 1190

\bibitem[\protect\citeauthoryear{{Dey} et~al.,}{{Dey}  et~al.}{2009}]{dey09}
{Dey} A. et~al., 2009, in {W.~Wang, Z.~Yang, Z.~Luo, \& Z.~Chen} ed.,
  Astronomical Society of the Pacific Conference Series Vol.~408 of
  Astronomical Society of the Pacific Conference Series, {The Pedigrees of DOGs
  (Dust-Obscured Galaxies)}.
pp 411--+

\bibitem[\protect\citeauthoryear{{Dey}, {Soifer}, {Desai}, {Brand}, {Le
  Floc'h}, {Brown}, {Jannuzi}, {Armus}, {Bussmann}, {Brodwin}, {Bian},
  {Eisenhardt}, {Higdon}, {Weedman} \& {Willner}}{{Dey} et~al.}{2008}]{dey08}
{Dey} A.,  {Soifer} B.~T.,  {Desai} V. et~al., 2008, \apj, 677, 943

\bibitem[\protect\citeauthoryear{{Di Matteo}, {Springel} \& {Hernquist}}{{Di
  Matteo} et~al.}{2005}]{dim05}
{Di Matteo} T.,  {Springel} V., {Hernquist} L.,  2005, \nat, 433, 604

\bibitem[\protect\citeauthoryear{{Dickinson}, {Papovich}, {Ferguson} \&
  {Budav{\'a}ri}}{{Dickinson} et~al.}{2003}]{dic03}
{Dickinson} M.,  {Papovich} C.,  {Ferguson} H.~C., {Budav{\'a}ri} T.,  2003,
  \apj, 587, 25

\bibitem[\protect\citeauthoryear{{Donley}, {Rieke}, {P{\'e}rez-Gonz{\'a}lez},
  {Rigby} \& {Alonso-Herrero}}{{Donley} et~al.}{2007}]{don07}
{Donley} J.~L.,  {Rieke} G.~H.,  {P{\'e}rez-Gonz{\'a}lez} P.~G.,  {Rigby}
  J.~R., {Alonso-Herrero} A.,  2007, \apj, 660, 167

\bibitem[\protect\citeauthoryear{{Dopita}, {Groves}, {Fischera}, {Sutherland},
  {Tuffs}, {Popescu}, {Kewley}, {Reuland} \& {Leitherer}}{{Dopita}
  et~al.}{2005}]{dop05}
{Dopita} M.~A.,  {Groves} B.~A.,  {Fischera} J. et~al., 2005, \apj, 619, 755

\bibitem[\protect\citeauthoryear{{Downes} \& {Solomon}}{{Downes} \&
  {Solomon}}{1998}]{dow98}
{Downes} D., {Solomon} P.~M.,  1998, \apj, 507, 615

\bibitem[\protect\citeauthoryear{{Draine} \& {Li}}{{Draine} \&
  {Li}}{2007}]{dra07}
{Draine} B.~T., {Li} A.,  2007, \apj, 657, 810

\bibitem[\protect\citeauthoryear{{Dwek}}{{Dwek}}{1998}]{dwe98}
{Dwek} E. 1998, \apj, 501, 643

\bibitem[\protect\citeauthoryear{{Elmegreen}}{{Elmegreen}}{2009}]{elm09c}
{Elmegreen} B.~G. 2009, ArXiv e-prints

\bibitem[\protect\citeauthoryear{{Elmegreen} \& {Burkert}}{{Elmegreen} \&
  {Burkert}}{2010}]{elm10}
{Elmegreen} B.~G., {Burkert} A.,  2010, \apj, 712, 294

\bibitem[\protect\citeauthoryear{{Elmegreen}, {Elmegreen}, {Marcus},
  {Shahinyan}, {Yau} \& {Petersen}}{{Elmegreen} et~al.}{2009}]{elm09a}
{Elmegreen} D.~M.,  {Elmegreen} B.~G.,  {Marcus} M.~T.,  {Shahinyan} K.,  {Yau}
  A., {Petersen} M.,  2009, \apj, 701, 306

\bibitem[\protect\citeauthoryear{{Elmegreen}, {Elmegreen}, {Ravindranath} \&
  {Coe}}{{Elmegreen} et~al.}{2007}]{elm07}
{Elmegreen} D.~M.,  {Elmegreen} B.~G.,  {Ravindranath} S., {Coe} D.~A.,  2007,
  \apj, 658, 763

\bibitem[\protect\citeauthoryear{{Fakhouri} \& {Ma}}{{Fakhouri} \&
  {Ma}}{2008}]{fak08}
{Fakhouri} O., {Ma} C.-P.,  2008, \mnras, 386, 577

\bibitem[\protect\citeauthoryear{{Farrah}, {Lonsdale}, {Weedman}, {Spoon},
  {Rowan-Robinson}, {Polletta}, {Oliver}, {Houck} \& {Smith}}{{Farrah}
  et~al.}{2008}]{far08}
{Farrah} D.,  {Lonsdale} C.~J.,  {Weedman} D.~W. et~al., 2008, \apj, 677, 957

\bibitem[\protect\citeauthoryear{{Fiore}, {Grazian}, {Santini}, {Puccetti},
  {Brusa}, {Feruglio}, {Fontana}, {Giallongo}, {Comastri}, {Gruppioni},
  {Pozzi}, {Zamorani} \& {Vignali}}{{Fiore} et~al.}{2008}]{fio08}
{Fiore} F.,  {Grazian} A.,  {Santini} P. et~al., 2008, \apj, 672, 94

\bibitem[\protect\citeauthoryear{{Fiore}, {Puccetti}, {Brusa}, {Salvato},
  {Zamorani}, {Aldcroft}, {Aussel}, {Brunner}, {Capak} \& {Cappelluti}}{{Fiore}
  et~al.}{2009}]{fio09}
{Fiore} F.,  {Puccetti} S.,  {Brusa} M. et~al., 2009, \apj, 693, 447

\bibitem[\protect\citeauthoryear{{F{\"o}rster Schreiber} et~al.,}{{F{\"o}rster
  Schreiber}  et~al.}{2009}]{for09}
{F{\"o}rster Schreiber} N.~M. et~al., 2009, \apj, 706, 1364

\bibitem[\protect\citeauthoryear{{Genzel} et~al.,}{{Genzel}
  et~al.}{2006}]{gen06}
{Genzel} R. et~al., 2006, \nat, 442, 786

\bibitem[\protect\citeauthoryear{{Georgantopoulos}, {Akylas}, {Georgakakis} \&
  {Rowan-Robinson}}{{Georgantopoulos} et~al.}{2009}]{geo09}
{Georgantopoulos} I.,  {Akylas} A.,  {Georgakakis} A., {Rowan-Robinson} M.,
  2009, ArXiv e-prints

\bibitem[\protect\citeauthoryear{{Greve} et~al.,}{{Greve}
  et~al.}{2005}]{gre05}
{Greve} T.~R. et~al., 2005, \mnras, 359, 1165

\bibitem[\protect\citeauthoryear{{Groves}, {Dopita}, {Sutherland}, {Kewley},
  {Fischera}, {Leitherer}, {Brandl} \& {van Breugel}}{{Groves}
  et~al.}{2008}]{gro08}
{Groves} B.,  {Dopita} M.~A.,  {Sutherland} R.~S.,  {Kewley} L.~J.,  {Fischera}
  J.,  {Leitherer} C.,  {Brandl} B., {van Breugel} W.,  2008, \apjs, 176, 438

\bibitem[\protect\citeauthoryear{{Hernquist}}{{Hernquist}}{1990}]{her90}
{Hernquist} L. 1990, \apj, 356, 359

\bibitem[\protect\citeauthoryear{{Hopkins}}{{Hopkins}}{2004}]{hop04}
{Hopkins} A.~M. 2004, \apj, 615, 209

\bibitem[\protect\citeauthoryear{{Hopkins}, {Bundy}, {Croton}, {Hernquist},
  {Keres}, {Khochfar}, {Stewart}, {Wetzel} \& {Younger}}{{Hopkins}
  et~al.}{2009}]{hop09d}
{Hopkins} P.~F.,  {Bundy} K.,  {Croton} D. et~al., 2009, ArXiv e-prints

\bibitem[\protect\citeauthoryear{{Hopkins} et~al.,}{{Hopkins}
  et~al.}{2005a}]{hop05b}
{Hopkins} P.~F. et~al., 2005a, \apjl, 625, L71

\bibitem[\protect\citeauthoryear{{Hopkins} et~al.,}{{Hopkins}
  et~al.}{2005b}]{hop05a}
{Hopkins} P.~F. et~al., 2005b, \apj, 630, 705

\bibitem[\protect\citeauthoryear{{Hopkins} et~al.,}{{Hopkins}
  et~al.}{2006}]{hop06}
{Hopkins} P.~F. et~al., 2006, \apjs, 163, 1

\bibitem[\protect\citeauthoryear{{Hopkins} et~al.,}{{Hopkins}
  et~al.}{2008}]{hop08a}
{Hopkins} P.~F. et~al., 2008, \apjs, 175, 356

\bibitem[\protect\citeauthoryear{{Hopkins} \& {Hernquist}}{{Hopkins} \&
  {Hernquist}}{2010}]{hop10b}
{Hopkins} P.~F., {Hernquist} L.,  2010, \mnras, 402, 985

\bibitem[\protect\citeauthoryear{{Hopkins}, {Hernquist}, {Cox}, {Robertson} \&
  {Krause}}{{Hopkins} et~al.}{2007}]{hop07b}
{Hopkins} P.~F.,  {Hernquist} L.,  {Cox} T.~J.,  {Robertson} B., {Krause} E.,
  2007, \apj, 669, 45

\bibitem[\protect\citeauthoryear{{Hopkins}, {Lidz}, {Hernquist}, {Coil},
  {Myers}, {Cox} \& {Spergel}}{{Hopkins} et~al.}{2007}]{hop07c}
{Hopkins} P.~F.,  {Lidz} A.,  {Hernquist} L.,  {Coil} A.~L.,  {Myers} A.~D.,
  {Cox} T.~J., {Spergel} D.~N.,  2007, \apj, 662, 110

\bibitem[\protect\citeauthoryear{{Hopkins}, {Richards} \&
  {Hernquist}}{{Hopkins} et~al.}{2007}]{hop07}
{Hopkins} P.~F.,  {Richards} G.~T., {Hernquist} L.,  2007, \apj, 654, 731

\bibitem[\protect\citeauthoryear{{Hopkins}, {Younger}, {Hayward}, {Narayanan}
  \& {Hernquist}}{{Hopkins} et~al.}{2010}]{hop10}
{Hopkins} P.~F.,  {Younger} J.~D.,  {Hayward} C.~C.,  {Narayanan} D.,
  {Hernquist} L.,  2010, \mnras, 402, 1693

\bibitem[\protect\citeauthoryear{{Houck}, {Soifer}, {Weedman}, {Higdon},
  {Higdon}, {Herter}, {Brown}, {Dey}, {Jannuzi}, {Le Floc'h}, {Rieke}, {Armus},
  {Charmandaris}, {Brandl} \& {Teplitz}}{{Houck} et~al.}{2005}]{hou05}
{Houck} J.~R.,  {Soifer} B.~T.,  {Weedman} D. et~al., 2005, \apjl, 622, L105

\bibitem[\protect\citeauthoryear{{Huang}, {Faber}, {Daddi}, {Laird}, {Lai},
  {Omont}, {Wu}, {Younger}, {Bundy} \& {Cattaneo}}{{Huang}
  et~al.}{2009}]{hua09}
{Huang} J.-S.,  {Faber} S.~M.,  {Daddi} E. et~al., 2009, \apj, 700, 183

\bibitem[\protect\citeauthoryear{{John}}{{John}}{1988}]{joh88}
{John} T.~L. 1988, \aap, 193, 189

\bibitem[\protect\citeauthoryear{{Jonsson}}{{Jonsson}}{2006}]{jon06a}
{Jonsson} P. 2006, \mnras, 372, 2

\bibitem[\protect\citeauthoryear{{Jonsson}, {Cox}, {Primack} \&
  {Somerville}}{{Jonsson} et~al.}{2006}]{jon06b}
{Jonsson} P.,  {Cox} T.~J.,  {Primack} J.~R., {Somerville} R.~S.,  2006, \apj,
  637, 255

\bibitem[\protect\citeauthoryear{{Jonsson}, {Groves} \& {Cox}}{{Jonsson}
  et~al.}{2010}]{jon10}
{Jonsson} P.,  {Groves} B.~A., {Cox} T.~J.,  2010, \mnras, pp 186--+

\bibitem[\protect\citeauthoryear{{Juvela}}{{Juvela}}{2005}]{juv05}
{Juvela} M. 2005, \aap, 440, 531

\bibitem[\protect\citeauthoryear{{Kennicutt} Jr.}{{Kennicutt}}{1998}]{ken98b}
{Kennicutt} Jr. R.~C. 1998, \apj, 498, 541

\bibitem[\protect\citeauthoryear{{Kere{\v s}}, {Katz}, {Fardal}, {Dav{\'e}} \&
  {Weinberg}}{{Kere{\v s}} et~al.}{2009}]{ker09}
{Kere{\v s}} D.,  {Katz} N.,  {Fardal} M.,  {Dav{\'e}} R., {Weinberg} D.~H.,
  2009, \mnras, 395, 160

\bibitem[\protect\citeauthoryear{{Kere{\v s}}, {Katz}, {Weinberg} \&
  {Dav{\'e}}}{{Kere{\v s}} et~al.}{2005}]{ker05}
{Kere{\v s}} D.,  {Katz} N.,  {Weinberg} D.~H., {Dav{\'e}} R.,  2005, \mnras,
  363, 2

\bibitem[\protect\citeauthoryear{{Kov{\'a}cs} et~al.,}{{Kov{\'a}cs}
  et~al.}{2006}]{kov06}
{Kov{\'a}cs} A. et~al., 2006, \apj, 650, 592

\bibitem[\protect\citeauthoryear{{Kov{\'a}cs}, {Omont}, {Beelen}, {Lonsdale},
  {Polletta}, {Fiolet}, {Greve}, {Borys}, {Cox}, {De Breuck}, {Dole}, {Dowell},
  {Farrah}, {Lagache}, {Menten}, {Bell} \& {Owen}}{{Kov{\'a}cs}
  et~al.}{2010}]{kov10}
{Kov{\'a}cs} A.,  {Omont} A.,  {Beelen} A. et~al., 2010, ArXiv e-prints

\bibitem[\protect\citeauthoryear{{Lacy}, {Storrie-Lombardi}, {Sajina},
  {Appleton}, {Armus}, {Chapman}, {Choi}, {Fadda}, {Fang} \& {Frayer}}{{Lacy}
  et~al.}{2004}]{lac04}
{Lacy} M.,  {Storrie-Lombardi} L.~J.,  {Sajina} A. et~al., 2004, \apjs, 154,
  166

\bibitem[\protect\citeauthoryear{{Le Floc'h}, {Papovich}, {Dole}, {Bell},
  {Lagache}, {Rieke}, {Egami}, {P{\'e}rez-Gonz{\'a}lez}, {Alonso-Herrero} \&
  {Rieke}}{{Le Floc'h} et~al.}{2005}]{lef05}
{Le Floc'h} E.,  {Papovich} C.,  {Dole} H. et~al., 2005, \apj, 632, 169

\bibitem[\protect\citeauthoryear{{Leitherer} et~al.,}{{Leitherer}
  et~al.}{1999}]{lei99}
{Leitherer} C. et~al., 1999, \apjs, 123, 3

\bibitem[\protect\citeauthoryear{{Li} \& {Draine}}{{Li} \&
  {Draine}}{2001}]{li01}
{Li} A., {Draine} B.~T.,  2001, \apj, 554, 778

\bibitem[\protect\citeauthoryear{{Li}, {Hernquist}, {Robertson}, {Cox},
  {Hopkins}, {Springel}, {Gao}, {Di Matteo}, {Zentner}, {Jenkins} \&
  {Yoshida}}{{Li} et~al.}{2007}]{li07}
{Li} Y.,  {Hernquist} L.,  {Robertson} B. et~al., 2007, \apj, 665, 187

\bibitem[\protect\citeauthoryear{{Lidz}, {Hopkins}, {Cox}, {Hernquist} \&
  {Robertson}}{{Lidz} et~al.}{2006}]{lid06}
{Lidz} A.,  {Hopkins} P.~F.,  {Cox} T.~J.,  {Hernquist} L., {Robertson} B.,
  2006, \apj, 641, 41

\bibitem[\protect\citeauthoryear{{Lonsdale}, {Polletta}, {Omont}, {Shupe},
  {Berta}, {Zylka}, {Siana}, {Lutz}, {Farrah} \& {Smith}}{{Lonsdale}
  et~al.}{2009}]{lon09}
{Lonsdale} C.~J.,  {Polletta} M.~d.~C.,  {Omont} A. et~al., 2009, \apj, 692,
  422

\bibitem[\protect\citeauthoryear{{Lotz}, {Jonsson}, {Cox} \& {Primack}}{{Lotz}
  et~al.}{2008}]{lot08}
{Lotz} J.~M.,  {Jonsson} P.,  {Cox} T.~J., {Primack} J.~R.,  2008, \mnras, 391,
  1137

\bibitem[\protect\citeauthoryear{{Melbourne}, {Bussman}, {Brand}, {Desai},
  {Armus}, {Dey}, {Jannuzi}, {Houck}, {Matthews} \& {Soifer}}{{Melbourne}
  et~al.}{2009}]{mel09}
{Melbourne} J.,  {Bussman} R.~S.,  {Brand} K. et~al., 2009, \aj, 137, 4854

\bibitem[\protect\citeauthoryear{{Melbourne}, {Desai}, {Armus}, {Dey}, {Brand},
  {Thompson}, {Soifer}, {Matthews}, {Jannuzi} \& {Houck}}{{Melbourne}
  et~al.}{2008}]{mel08}
{Melbourne} J.,  {Desai} V.,  {Armus} L. et~al., 2008, \aj, 136, 1110

\bibitem[\protect\citeauthoryear{{Men{\'e}ndez-Delmestre}, {Blain},
  {Alexander}, {Smail}, {Armus}, {Chapman}, {Frayer}, {Ivison} \&
  {Teplitz}}{{Men{\'e}ndez-Delmestre} et~al.}{2007}]{men07}
{Men{\'e}ndez-Delmestre} K.,  {Blain} A.~W.,  {Alexander} D.~M. et~al., 2007,
  \apjl, 655, L65

\bibitem[\protect\citeauthoryear{{Men{\'e}ndez-Delmestre}, {Blain}, {Smail},
  {Alexander}, {Chapman}, {Armus}, {Frayer}, {Ivison} \&
  {Teplitz}}{{Men{\'e}ndez-Delmestre} et~al.}{2009}]{men09}
{Men{\'e}ndez-Delmestre} K.,  {Blain} A.~W.,  {Smail} I. et~al., 2009, \apj,
  699, 667

\bibitem[\protect\citeauthoryear{{Micha{\l}owski}, {Hjorth} \&
  {Watson}}{{Micha{\l}owski} et~al.}{2009}]{mic09}
{Micha{\l}owski} M.~J.,  {Hjorth} J., {Watson} D.,  2009, ArXiv e-prints

\bibitem[\protect\citeauthoryear{{Mihos} \& {Hernquist}}{{Mihos} \&
  {Hernquist}}{1994}]{mih94a}
{Mihos} J.~C., {Hernquist} L.,  1994, \apjl, 431, L9

\bibitem[\protect\citeauthoryear{{Mihos} \& {Hernquist}}{{Mihos} \&
  {Hernquist}}{1996}]{mih96}
{Mihos} J.~C., {Hernquist} L.,  1996, \apj, 464, 641

\bibitem[\protect\citeauthoryear{{Mo}, {Mao} \& {White}}{{Mo}
  et~al.}{1998}]{mo98}
{Mo} H.~J.,  {Mao} S., {White} S.~D.~M.,  1998, \mnras, 295, 319

\bibitem[\protect\citeauthoryear{{Murphy}, {Chary}, {Alexander}, {Dickinson},
  {Magnelli}, {Morrison}, {Pope} \& {Teplitz}}{{Murphy} et~al.}{2009}]{mur09}
{Murphy} E.~J.,  {Chary} R.-R.,  {Alexander} D.~M.,  {Dickinson} M.,
  {Magnelli} B.,  {Morrison} G.,  {Pope} A., {Teplitz} H.~I.,  2009, \apj, 698,
  1380

\bibitem[\protect\citeauthoryear{{Narayanan}, {Cox}, {Hayward}, {Younger} \&
  {Hernquist}}{{Narayanan} et~al.}{2009}]{nar09b}
{Narayanan} D.,  {Cox} T.~J.,  {Hayward} C.~C.,  {Younger} J.~D., {Hernquist}
  L.,  2009, \mnras, 400, 1919

\bibitem[\protect\citeauthoryear{{Narayanan}, {Cox}, {Kelly}, {Dav{\'e}},
  {Hernquist}, {Di Matteo}, {Hopkins}, {Kulesa}, {Robertson} \&
  {Walker}}{{Narayanan} et~al.}{2008}]{nar08a}
{Narayanan} D.,  {Cox} T.~J.,  {Kelly} B. et~al., 2008, \apjs, 176, 331

\bibitem[\protect\citeauthoryear{{Narayanan}, {Hayward}, {Cox}, {Hernquist},
  {Jonsson}, {Younger} \& {Groves}}{{Narayanan} et~al.}{2009}]{nar09}
{Narayanan} D.,  {Hayward} C.~C.,  {Cox} T.~J.,  {Hernquist} L.,  {Jonsson} P.,
   {Younger} J.~D., {Groves} B.,  2009, MNRAS in press: arXiv/0904.0004

\bibitem[\protect\citeauthoryear{{Narayanan}, {Hayward}, {Cox}, {Hernquist},
  {Jonsson}, {Younger} \& {Groves}}{{Narayanan} et~al.}{2010}]{nar10a}
{Narayanan} D.,  {Hayward} C.~C.,  {Cox} T.~J.,  {Hernquist} L.,  {Jonsson} P.,
   {Younger} J.~D., {Groves} B.,  2010, \mnras, 401, 1613

\bibitem[\protect\citeauthoryear{{Narayanan}, {Kulesa}, {Boss} \&
  {Walker}}{{Narayanan} et~al.}{2006}]{nar06b}
{Narayanan} D.,  {Kulesa} C.~A.,  {Boss} A., {Walker} C.~K.,  2006, \apj, 647,
  1426

\bibitem[\protect\citeauthoryear{{Narayanan}, {Li}, {Cox}, {Hernquist},
  {Hopkins}, {Chakrabarti}, {Dav{\'e}}, {Di Matteo}, {Gao}, {Kulesa},
  {Robertson} \& {Walker}}{{Narayanan} et~al.}{2008}]{nar08c}
{Narayanan} D.,  {Li} Y.,  {Cox} T.~J. et~al., 2008, \apjs, 174, 13

\bibitem[\protect\citeauthoryear{{P{\'e}rez-Gonz{\'a}lez}, {Rieke}, {Egami},
  {Alonso-Herrero}, {Dole}, {Papovich}, {Blaylock}, {Jones}, {Rieke}, {Rigby},
  {Barmby}, {Fazio}, {Huang} \& {Martin}}{{P{\'e}rez-Gonz{\'a}lez}
  et~al.}{2005}]{per05}
{P{\'e}rez-Gonz{\'a}lez} P.~G.,  {Rieke} G.~H.,  {Egami} E. et~al., 2005, \apj,
  630, 82

\bibitem[\protect\citeauthoryear{{Pope}, {Bussmann}, {Dey}, {Meger},
  {Alexander}, {Brodwin}, {Chary}, {Dickinson}, {Frayer}, {Greve}, {Huynh},
  {Lin}, {Morrison}, {Scott} \& {Yan}}{{Pope} et~al.}{2008}]{pop08b}
{Pope} A.,  {Bussmann} R.~S.,  {Dey} A. et~al., 2008, \apj, 689, 127

\bibitem[\protect\citeauthoryear{{Pope}, {Chary}, {Alexander}, {Armus},
  {Dickinson}, {Elbaz}, {Frayer}, {Scott} \& {Teplitz}}{{Pope}
  et~al.}{2008}]{pop08}
{Pope} A.,  {Chary} R.-R.,  {Alexander} D.~M. et~al., 2008, \apj, 675, 1171

\bibitem[\protect\citeauthoryear{{Reddy}, {Steidel}, {Pettini}, {Adelberger},
  {Shapley}, {Erb} \& {Dickinson}}{{Reddy} et~al.}{2008}]{red08}
{Reddy} N.~A.,  {Steidel} C.~C.,  {Pettini} M.,  {Adelberger} K.~L.,  {Shapley}
  A.~E.,  {Erb} D.~K., {Dickinson} M.,  2008, \apjs, 175, 48

\bibitem[\protect\citeauthoryear{{Richards}, {Lacy}, {Storrie-Lombardi},
  {Hall}, {Gallagher}, {Hines}, {Fan}, {Papovich}, {Vanden Berk}, {Trammell},
  {Schneider}, {Vestergaard}, {York}, {Jester}, {Anderson}, {Budav{\'a}ri} \&
  {Szalay}}{{Richards} et~al.}{2006}]{ric06}
{Richards} G.~T.,  {Lacy} M.,  {Storrie-Lombardi} L.~J. et~al., 2006, \apjs,
  166, 470

\bibitem[\protect\citeauthoryear{{Rieke}, {Alonso-Herrero}, {Weiner},
  {P{\'e}rez-Gonz{\'a}lez}, {Blaylock}, {Donley} \& {Marcillac}}{{Rieke}
  et~al.}{2009}]{rie09}
{Rieke} G.~H.,  {Alonso-Herrero} A.,  {Weiner} B.~J.,  {P{\'e}rez-Gonz{\'a}lez}
  P.~G.,  {Blaylock} M.,  {Donley} J.~L., {Marcillac} D.,  2009, \apj, 692, 556

\bibitem[\protect\citeauthoryear{{Rigby}, {Rieke}, {Maiolino}, {Gilli},
  {Papovich}, {P{\'e}rez-Gonz{\'a}lez}, {Alonso-Herrero}, {Le Floc'h},
  {Engelbracht}, {Gordon}, {Hines}, {Hinz}, {Morrison}, {Muzerolle}, {Rieke} \&
  {Su}}{{Rigby} et~al.}{2004}]{rig04}
{Rigby} J.~R.,  {Rieke} G.~H.,  {Maiolino} R. et~al., 2004, \apjs, 154, 160

\bibitem[\protect\citeauthoryear{{Robertson}, {Hernquist}, {Cox}, {Di Matteo},
  {Hopkins}, {Martini} \& {Springel}}{{Robertson} et~al.}{2006}]{rob06b}
{Robertson} B.,  {Hernquist} L.,  {Cox} T.~J.,  {Di Matteo} T.,  {Hopkins}
  P.~F.,  {Martini} P., {Springel} V.,  2006, \apj, 641, 90

\bibitem[\protect\citeauthoryear{{Rosolowsky}}{{Rosolowsky}}{2005}]{ros05}
{Rosolowsky} E. 2005, \pasp, 117, 1403

\bibitem[\protect\citeauthoryear{{Rudnick}, {Labb{\'e}}, {F{\"o}rster
  Schreiber}, {Wuyts}, {Franx}, {Finlator}, {Kriek}, {Moorwood}, {Rix},
  {R{\"o}ttgering}, {Trujillo}, {van der Wel}, {van der Werf} \& {van
  Dokkum}}{{Rudnick} et~al.}{2006}]{rud06}
{Rudnick} G.,  {Labb{\'e}} I.,  {F{\"o}rster Schreiber} N.~M. et~al., 2006,
  \apj, 650, 624

\bibitem[\protect\citeauthoryear{{Sacchi}, {La Franca}, {Feruglio}, {Fiore},
  {Puccetti}, {Cocchia}, {Berta}, {Brusa}, {Cimatti} \& {Comastri}}{{Sacchi}
  et~al.}{2009}]{sac09}
{Sacchi} N.,  {La Franca} F.,  {Feruglio} C. et~al., 2009, ArXiv e-prints

\bibitem[\protect\citeauthoryear{{Sajina}, {Yan}, {Armus}, {Choi}, {Fadda},
  {Helou} \& {Spoon}}{{Sajina} et~al.}{2007}]{saj07}
{Sajina} A.,  {Yan} L.,  {Armus} L.,  {Choi} P.,  {Fadda} D.,  {Helou} G.,
  {Spoon} H.,  2007, \apj, 664, 713

\bibitem[\protect\citeauthoryear{{Sakamoto} et~al.,}{{Sakamoto}
  et~al.}{1999}]{sak99}
{Sakamoto} K. et~al., 1999, \apj, 514, 68

\bibitem[\protect\citeauthoryear{{Sanders} \& {Mirabel}}{{Sanders} \&
  {Mirabel}}{1996}]{san96}
{Sanders} D.~B., {Mirabel} I.~F.,  1996, \araa, 34, 749

\bibitem[\protect\citeauthoryear{{Sanders}, {Soifer}, {Elias}, {Madore},
  {Matthews}, {Neugebauer} \& {Scoville}}{{Sanders} et~al.}{1988}]{san88}
{Sanders} D.~B.,  {Soifer} B.~T.,  {Elias} J.~H.,  {Madore} B.~F.,  {Matthews}
  K.,  {Neugebauer} G., {Scoville} N.~Z.,  1988, \apj, 325, 74

\bibitem[\protect\citeauthoryear{{Shaver}, {Wall}, {Kellermann}, {Jackson} \&
  {Hawkins}}{{Shaver} et~al.}{1996}]{sha96}
{Shaver} P.~A.,  {Wall} J.~V.,  {Kellermann} K.~I.,  {Jackson} C.~A., {Hawkins}
  M.~R.~S.,  1996, \nat, 384, 439

\bibitem[\protect\citeauthoryear{{Shen}, {Strauss}, {Oguri}, {Hennawi}, {Fan},
  {Richards}, {Hall}, {Gunn}, {Schneider}, {Szalay}, {Thakar}, {Vanden Berk},
  {Anderson}, {Bahcall}, {Connolly} \& {Knapp}}{{Shen} et~al.}{2007}]{she07}
{Shen} Y.,  {Strauss} M.~A.,  {Oguri} M. et~al., 2007, \aj, 133, 2222

\bibitem[\protect\citeauthoryear{{Shi}, {Rieke}, {Lotz} \&
  {Perez-Gonzalez}}{{Shi} et~al.}{2009}]{shi09}
{Shi} Y.,  {Rieke} G.,  {Lotz} J., {Perez-Gonzalez} P.~G.,  2009, \apj, 697,
  1764

\bibitem[\protect\citeauthoryear{{Simpson} \& {Eisenhardt}}{{Simpson} \&
  {Eisenhardt}}{1999}]{sim99}
{Simpson} C., {Eisenhardt} P.,  1999, \pasp, 111, 691

\bibitem[\protect\citeauthoryear{{Soifer}, {Helou} \& {Werner}}{{Soifer}
  et~al.}{2008}]{soi08}
{Soifer} B.~T.,  {Helou} G., {Werner} M.,  2008, \araa, 46, 201

\bibitem[\protect\citeauthoryear{{Solomon}, {Rivolo}, {Barrett} \&
  {Yahil}}{{Solomon} et~al.}{1987}]{sol87}
{Solomon} P.~M.,  {Rivolo} A.~R.,  {Barrett} J., {Yahil} A.,  1987, \apj, 319,
  730

\bibitem[\protect\citeauthoryear{{Springel}}{{Springel}}{2005}]{spr05b}
{Springel} V. 2005, \mnras, 364, 1105

\bibitem[\protect\citeauthoryear{{Springel}, {Di Matteo} \&
  {Hernquist}}{{Springel} et~al.}{2005a}]{spr05c}
{Springel} V.,  {Di Matteo} T., {Hernquist} L.,  2005a, \apjl, 620, L79

\bibitem[\protect\citeauthoryear{{Springel}, {Di Matteo} \&
  {Hernquist}}{{Springel} et~al.}{2005b}]{spr05a}
{Springel} V.,  {Di Matteo} T., {Hernquist} L.,  2005b, \mnras, 361, 776

\bibitem[\protect\citeauthoryear{{Springel} \& {Hernquist}}{{Springel} \&
  {Hernquist}}{2002}]{spr02}
{Springel} V., {Hernquist} L.,  2002, \mnras, 333, 649

\bibitem[\protect\citeauthoryear{{Springel} \& {Hernquist}}{{Springel} \&
  {Hernquist}}{2003}]{spr03a}
{Springel} V., {Hernquist} L.,  2003, \mnras, 339, 289

\bibitem[\protect\citeauthoryear{{Swinbank} et~al.,}{{Swinbank}
  et~al.}{2004}]{swi04}
{Swinbank} A.~M. et~al., 2004, \apj, 617, 64

\bibitem[\protect\citeauthoryear{{Swinbank} et~al.,}{{Swinbank}
  et~al.}{2008}]{swi08}
{Swinbank} A.~M. et~al., 2008, \mnras, 391, 420

\bibitem[\protect\citeauthoryear{{Tacconi} et~al.,}{{Tacconi}
  et~al.}{2006}]{tac06}
{Tacconi} L.~J. et~al., 2006, \apj, 640, 228

\bibitem[\protect\citeauthoryear{{Tacconi} et~al.,}{{Tacconi}
  et~al.}{2008}]{tac08}
{Tacconi} L.~J. et~al., 2008, \apj, 680, 246

\bibitem[\protect\citeauthoryear{{Tacconi} et~al.,}{{Tacconi}
  et~al.}{2010}]{tac10}
{Tacconi} L.~J. et~al., 2010, \nat, 463, 781

\bibitem[\protect\citeauthoryear{{Tyler}, {Floc'h}, {Rieke}, {Dey}, {Desai},
  {Brand}, {Borys}, {Jannuzi}, {Armus}, {Dole}, {Papovich}, {Brown},
  {Blaylock}, {Higdon}, {Higdon}, {Charmandaris}, {Ashby} \& {Smith}}{{Tyler}
  et~al.}{2009}]{tyl09}
{Tyler} K.~D.,  {Floc'h} E.~L.,  {Rieke} G.~H. et~al., 2009, \apj, 691, 1846

\bibitem[\protect\citeauthoryear{{Valiante}, {Lutz}, {Sturm}, {Genzel},
  {Tacconi}, {Lehnert} \& {Baker}}{{Valiante} et~al.}{2007}]{val07}
{Valiante} E.,  {Lutz} D.,  {Sturm} E.,  {Genzel} R.,  {Tacconi} L.~J.,
  {Lehnert} M.~D., {Baker} A.~J.,  2007, \apj, 660, 1060

\bibitem[\protect\citeauthoryear{{V{\'a}zquez} \& {Leitherer}}{{V{\'a}zquez} \&
  {Leitherer}}{2005}]{vaz05}
{V{\'a}zquez} G.~A., {Leitherer} C.,  2005, \apj, 621, 695

\bibitem[\protect\citeauthoryear{{Vladilo}}{{Vladilo}}{1998}]{vla98}
{Vladilo} G. 1998, \apj, 493, 583

\bibitem[\protect\citeauthoryear{{Weedman} et~al.,}{{Weedman}
  et~al.}{2006a}]{wee06b}
{Weedman} D. et~al., 2006a, \apj, 653, 101

\bibitem[\protect\citeauthoryear{{Weedman} et~al.,}{{Weedman}
  et~al.}{2006b}]{wee06a}
{Weedman} D.~W. et~al., 2006b, \apj, 651, 101

\bibitem[\protect\citeauthoryear{{Weingartner} \& {Draine}}{{Weingartner} \&
  {Draine}}{2001}]{wei01}
{Weingartner} J.~C., {Draine} B.~T.,  2001, \apj, 548, 296

\bibitem[\protect\citeauthoryear{{Yan}, {Chary}, {Armus}, {Teplitz}, {Helou},
  {Frayer}, {Fadda}, {Surace} \& {Choi}}{{Yan} et~al.}{2005}]{yan05}
{Yan} L.,  {Chary} R.,  {Armus} L. et~al., 2005, \apj, 628, 604

\bibitem[\protect\citeauthoryear{{Yan}, {Sajina}, {Fadda}, {Choi}, {Armus},
  {Helou}, {Teplitz}, {Frayer} \& {Surace}}{{Yan} et~al.}{2007}]{yan07}
{Yan} L.,  {Sajina} A.,  {Fadda} D. et~al., 2007, \apj, 658, 778

\bibitem[\protect\citeauthoryear{{Yan}, {Tacconi}, {Fiolet}, {Sajina}, {Omont},
  {Lutz}, {Zamojski}, {Neri}, {Cox} \& {Dasyra}}{{Yan} et~al.}{2010}]{yan10}
{Yan} L.,  {Tacconi} L.~J.,  {Fiolet} N. et~al., 2010, \apj, 714, 100

\bibitem[\protect\citeauthoryear{{Younger} et~al.,}{{Younger}
  et~al.}{2008a}]{you08b}
{Younger} J.~D. et~al., 2008a, \apj, 688, 59

\bibitem[\protect\citeauthoryear{{Younger} et~al.,}{{Younger}
  et~al.}{2008b}]{you08a}
{Younger} J.~D. et~al., 2008b, \apj, 686, 815

\bibitem[\protect\citeauthoryear{{Younger}, {Hayward}, {Narayanan}, {Cox},
  {Hernquist} \& {Jonsson}}{{Younger} et~al.}{2009}]{you09}
{Younger} J.~D.,  {Hayward} C.~C.,  {Narayanan} D.,  {Cox} T.~J.,  {Hernquist}
  L., {Jonsson} P.,  2009, \mnras, 396, L66

\bibitem[\protect\citeauthoryear{{Younger}, {Omont}, {Fiolet}, {Huang},
  {Fazio}, {Lai}, {Polletta}, {Rigopoulou} \& {Zylka}}{{Younger}
  et~al.}{2009}]{you09b}
{Younger} J.~D.,  {Omont} A.,  {Fiolet} N. et~al., 2009, \mnras, 394, 1685

\end{thebibliography}

\end{document}